\documentclass[prb,longbibliography,twocolumn,final,superscriptaddress,english,showpacs]{revtex4-1} 
\usepackage[english]{babel}
\usepackage{xcolor} 
\usepackage{graphicx} 
\usepackage{amsmath} 
\usepackage{amssymb}

\definecolor{darkblue}{rgb}{0.1,0.2,0.6} \definecolor{darkred}{rgb}{0.8,0.1,0.2}
\usepackage[colorlinks,citecolor=darkblue,linkcolor=darkred,urlcolor=darkblue]{hyperref}
\usepackage[all]{hypcap} 

\makeatletter \adddialect\l@English\l@english \makeatother

\newcommand{\bra}[1]{\langle\,#1\,|} 
\newcommand{\ket}[1]{|#1\rangle}

\renewcommand{\vec}[1]{\boldsymbol{#1}}

\newcommand{\ie}{\textit{i.e.} } 
\newcommand{\eg}{\textit{e.g.} }

\newcommand{\tr}{\mathrm{Tr}}

\begin{document}
\title{Long tail distributions near the many body localization transition} 
\author{David J. Luitz}
\affiliation{Institute for Condensed Matter Theory and Department of Physics, University of Illinois at Urbana-Champaign, Urbana, IL 61801, USA}
\affiliation{Laboratoire de Physique Th\'eorique, Universit\'e de Toulouse, CNRS, UPS, France } 
\email{dluitz@illinois.edu}
\date{January 18, 2016}

\begin{abstract} 

    The random field $S=\frac 1 2$ Heisenberg chain exhibits a dynamical many body localization
    transition at a critical disorder strength, which depends on the energy density. At weak
    disorder, the eigenstate thermalization hypothesis (ETH) is fulfilled on average, making local
    observables smooth functions of energy, whose eigenstate-to-eigenstate fluctuations decrease
    exponentially with system size.

    We demonstrate the validity of ETH in the thermal phase as well as its breakdown in the
    localized phase and show that rare states exist which do not strictly follow ETH, 
    becoming more frequent closer to the transition. 
    Similarly, the probability distribution of the entanglement entropy at
    intermediate disorder develops long tails all the way down to zero entanglement. We propose that
    these low entanglement tails stem from localized regions at the subsystem boundaries which were
    recently discussed as a possible mechanism for subdiffusive transport in the ergodic phase.
\end{abstract} \pacs{75.10.Pq,03.65.Ud,71.30.+h}

\maketitle

\section{Introduction}

The eigenstate thermalization hypothesis
(ETH)\cite{deutsch_quantum_1991,srednicki_chaos_1994,dalessio_quantum_2015} is a mechanism for the
thermalization of generic closed quantum systems that has been numerically verified in many models
\cite{rigol_thermalization_2008,rigol_breakdown_2009,beugeling_finite-size_2014,steinigeweg_eigenstate_2013} and is also
argued to be valid for typical states of integrable systems, where the approach to the thermodynamic
limit is dramatically slower\cite{alba_eigenstate_2015}.  If it is fulfilled, it ensures that the
infinite time limit of local observables of any initial state will be equal to the canonical
expectation value at a temperature that is fixed by the energy of the initial state, implying that
eigenstates whose energies are close to each other have very similar properties and local operators
yield identical results in the thermodynamic limit. Consequently, for very large systems, local
observables vary smoothly with energy and microcanonical averages are well defined, corresponding to
the canonical expectation values. In systems of finite size, local observables from eigenstates in a
microcanonical energy window are distributed according to a normal distribution with a variance that
decreases as the square root of the dimension of the Hilbert
space\cite{ikeda_eigenstate_2011,rigol_alternatives_2012,beugeling_finite-size_2014,ikeda_how_2015},
due to the exponentially growing density of states with system size.

\begin{figure}[t]
    \centering
    \includegraphics{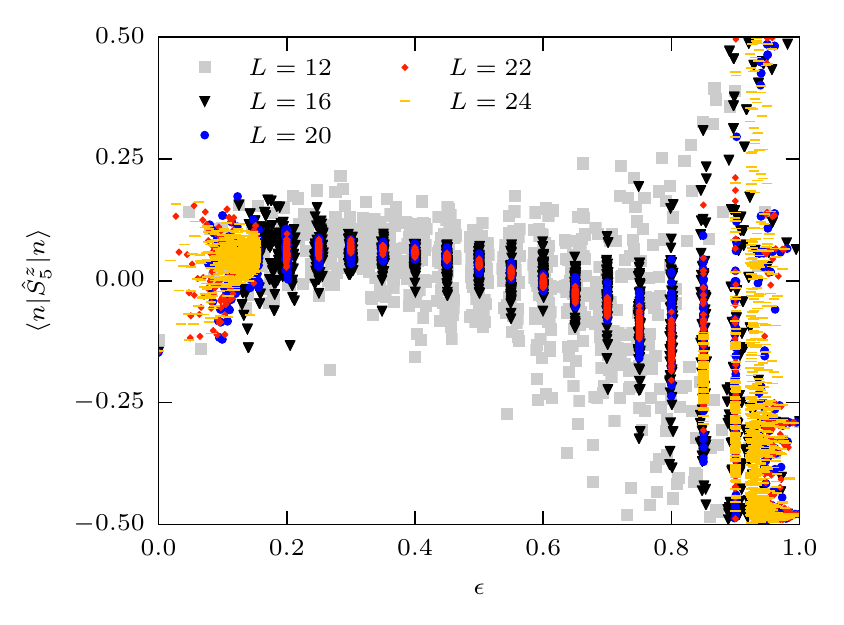}
    \vspace{-0.8cm}
    \caption{Local magnetization for site 5 in the chain as a function of energy. Here, we have
        generated one disorder configuration at $h=1.0$ for the largest system size $L_\text{max} =
        24$, smaller systems are cut out of this configuration for comparable results. The spread of the
    results in the middle of the spectrum reduces significantly with system size, thus leading to a
    \emph{smooth function of energy} in the thermodynamic limit. Clearly, at the high end of the
spectrum, this is no longer true and the variance of the local magnetization does not reduce with
system size. This is a signal for \emph{many body localization}, which is expected at $\epsilon
\gtrsim 0.9$ at $h=1$ in accord with Ref. \onlinecite{luitz_many-body_2015} and thus shows the
\emph{many body mobility edge}}
    \label{fig:obsvsenergy}
\end{figure}

This behavior is in stark contrast with many body localized (MBL) systems, where ETH breaks
down\cite{khatami_quantum_2012,mondaini_many-body_2015,nandkishore_many-body_2015} and
eigenstate expectation values of local operators in a very small energy window have a large 
(not decreasing with system size) variance. 
As a result, MBL systems do not thermalize and cannot act as their own heat
bath, which is the case if ETH is true. This can be seen in the area law behavior of the
entanglement entropy of eigenstates at arbitrary energy, which does not match the thermodynamic
volume law scaling that is necessary for ETH eigenstates.

Systems with a many body localization (MBL) transition from a thermal to a localized phase therefore
radically change their behavior at the critical point, where the validity of ETH breaks down,
leading to the characteristic behavior of eigenstates in the ETH (MBL) phase, which have volume
(area) law
entanglement\cite{bauer_area_2013,kjall_many-body_2014,grover_certain_2014,luitz_many-body_2015,bar_lev_absence_2015,vosk_theory_2015,chen_many-body_2015},
a vanishing (constant) variance of adjacent eigenstate expectation values of local operators with
system size\cite{pal_many-body_2010} and Wigner-Dyson (Poisson) level spacing
statistics\cite{georgeot_integrability_1998,pal_many-body_2010,luitz_many-body_2015}. Even more
interestingly, the position of the critical point depends on energy, forming a mobility edge
\cite{mondragon-shem_many-body_2015,luitz_many-body_2015,bera_many-body_2015,laumann_many-body_2014,serbyn_criterion_2015},
below which eigenstates are localized, while states at higher energy still fulfill the ETH.

The above discussion suggests that the ETH phase is in fact homogeneous up to the critical point and
can be perfectly described by random matrix ensembles\footnote{Except for the region very close to
the clean limit at $h=0$, where the system becomes integrable.}. Recent studies found, however, that
this can not be entirely true, as transport seems to be
subdiffusive\cite{bar_lev_absence_2015,vosk_theory_2015,agarwal_anomalous_2015,lerose_coexistence_2015,luitz_extended_2016,lerose_coexistence_2015}
in a large region of the ETH phase, with transport exponents that vary continuously with disorder
strength. In fact, renormalization group studies\cite{vosk_theory_2015,potter_universal_2015}
demonstrate that the anomalous transport is created by rare insulating regions which act as
bottlenecks for the transport and lead to Griffiths
physics\cite{griffiths_nonanalytic_1969,vojta_quantum_2010}. It is argued, however, that the
subdiffusive region is still thermalizing.

While a large part of the present numerical work on MBL has been devoted to single eigenstate
properties, the signature of subdiffusion is most obvious in the dynamical properties, for instance
the time evolution of observables after a quench. Furthermore, the observation of the growth of
the entanglement entropy with time, when starting from a product state shows the intriguing property
of a logarithmic
growth\cite{chiara_entanglement_2006,znidaric_many-body_2008,bardarson_unbounded_2012,luitz_extended_2016}
in the MBL phase, being now clearly understood in terms of the effective l-bit
Hamiltonian\cite{serbyn_local_2013,huse_phenomenology_2014,imbrie_many-body_2014,nandkishore_many-body_2015}
through exponentially decaying interaction terms with distance. In the ETH phase, one observes a
power law growth of the entanglement entropy with time, where the exponent varies continuously with
disorder strength\cite{luitz_extended_2016,vosk_theory_2015,potter_universal_2015}, in agreement
with the picture of transport bottlenecks.

This article presents a detailed analysis of the domain of validity of the ETH and the probability
distributions of the entanglement entropy as well as of eigenstate fluctuations of the local
magnetization. Our data is consistent the picture of the Griffiths phase at weak disorder, leading to
subdiffusive transport and slow entanglement dynamics, while ETH is still valid on average. We find
that the probability distributions of eigenstate-to-eigenstate fluctuations of local observables
show pronounced tails, deviating strongly from the normal distribution, at intermediate disorder
strength. At the same time, the probability distribution of the entanglement entropy shows low entanglement tails
that we argue to be connected to bottlenecks of transport, which can be observed in the spatial
entanglement structure.

\section{Model and Method}

  In this work, we focus on the zero magnetization sector of the periodic quantum Heisenberg chain
  subject to a random magnetic field, given by the Hamiltonian
  \begin{equation}
      H=\sum_{i} \vec{S}_i \cdot \vec{S}_{i+1} + h_i S_i^z, \quad h_{i} \in [-h,h],
      \label{eq:hamiltonian}
  \end{equation}
  where the local fields $h_i$ are drawn from a box distribution of width $2h$, and $h$ signifies
  the disorder strength.

  This model has been studied intensively in the context of many body localization, a dynamical
  quantum phase transition occurring in high energy eigenstates at a critical disorder strength
  $h_c$, which has been shown to depend on the energy density (with respect to the bandwidth)
  $\epsilon=(E-E_\text{min})/(E_\text{max}-E_\text{min})$ of the eigenstate
  \cite{pal_many-body_2010,vosk_theory_2015,mondragon-shem_many-body_2015,luca_ergodicity_2013,luitz_many-body_2015} (with the eigenenergy
  $E$, groundstate energy $E_\text{min}$ and antigroundstate energy $E_\text{max}$ of the sample).

  The transition manifests itself in the most striking difference of eigenstates in the two phases,
  thermalizing in the extended phase and breaking ETH in the localized phase. Here, we will study
  the validity of the ETH in the extended phase in detail, focussing on the scaling of deviations
  from ETH and the probability distributions of local operators at fixed energy.

  As explained in a previous work\cite{luitz_many-body_2015}, we calculate eigenstates corresponding
  to eigenenergies closest to a target energy density $\sigma$ for a large number of disorder
  configurations and different disorder strengths as well as system sizes, using a parallelized
  version of the shift invert technique. Typically, we calculate the $\gtrsim 50$ eigenpairs, whose
  energy density $\epsilon$  is closest to the target energy density $\sigma$. We will restrict most
  of the discussion in this work to the center of the band ($\epsilon=0.5$) unless stated
  differently. Our results are based on at least $10^3$ disorder realizations per system size and
  value of the disorder strength, except for $L=22$, where we could only afford to perfom
  calculations for $\gtrsim 100$ realizations.
  While the shift invert method works extremely well to access interior eigenpairs, the need for an
  accurate solution of a linear problem at each Lanzcos step makes it prohibitively expensive for
  larger systems. We find that coming from the edges of the spectrum, one can successfully determine
  a relatively large number of eigenpairs using a sophisticated deflation technique given by the
  Krylov-Schur algorithm\cite{stewart_krylov--schur_2002}, which was used for $L=24$ in Fig.
  \ref{fig:obsvsenergy}. For larger systems, the accessible eigenpairs will, however, lie
  increasingly close to $\epsilon=0$ and $\epsilon=1$.

  For the study of the ETH, it is crucial to compare eigenstate expectation values of the same
  operator in two eigenstates whose corresponding eigenenergies are closest to each other. We shall
  therefore call eigenstates with neighboring eigenenergies \emph{adjacent eigenstates}.

\begin{figure}[h]
    \centering
    \includegraphics{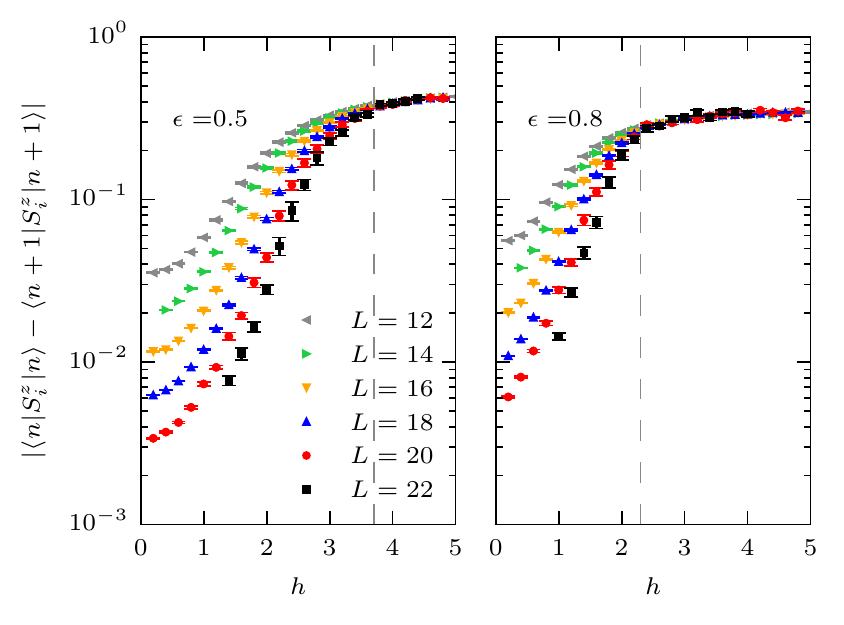}
\caption{Disorder averaged difference of local magnetizations as a function of disorder strength in
the middle (left) and upper part (right) of the spectrum. For weak disorder, eigenstates thermalize
and the local magnetization of adjacent states becomes identical in the thermodynamic limit, thus
yielding a smooth function of energy. Here, the mobility edge for the many body localization
transition is also visible as the region, in which adjacent states yield very different results --
the MBL phase -- is much larger in the upper part of the spectrum at energy density $\epsilon =
0.8$. The location of  the previously estimated\cite{luitz_many-body_2015} critical disorder
strength $h_c\approx3.7$ at $\epsilon=0.5$ and $h_c\approx2.3$ at $\epsilon=0.8$ is given by the
dashed line.}
    \label{fig:localmag}
\end{figure}

\begin{figure}[h]
    \centering
    \includegraphics{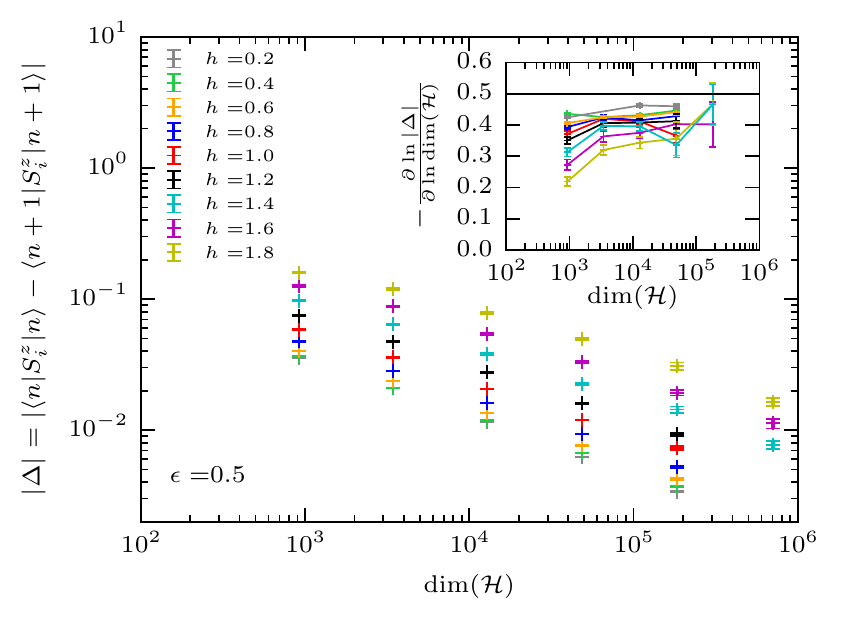}
    \caption{Finite size scaling of the disorder averaged difference of local magnetizations
        $\bra{n} S_z^i \ket{n}$ of adjacent eigenstates as a function of the dimension of the
        Hilbert space $\text{dim}(\mathcal{H})$. The inset displays the (discrete) logarithmic
    derivative, yielding the exponent of the power law in the thermodynamic limit, which slowly
    approaches $0.5$. The data presented here includes only states from the middle of the spectrum ($\epsilon=0.5$).} 
    \label{fig:diff_scaling}
\end{figure}
\section{Local magnetization}

Let us consider the simplest local operator to check the validity of the ETH: the local
magnetization $\hat{S}_i^z$ on site $i$ of the system. The qualitative behavior of this quantity as
a function of energy at intermediate disorder strength ($h=1.0$) is already clear from Fig.
\ref{fig:obsvsenergy}, where we show single eigenstate expectation values of $S_5^z$ at their
corresponding energy density $\epsilon$ for different system sizes. Note that the disorder
configuration of smaller system sizes is obtained as a subset of the larger sizes for comparability.
In the middle of the spectrum, the data corresponds to $\gtrsim 50$ eigenstates obtained by a
shift-invert technique, while for the largest system size $L=24$, we use a deflation technique for
$\lesssim2000$ states at the extreme ends of the spectrum, which successfully works down to
$\epsilon=0.85$ for this size.

In Fig. \ref{fig:obsvsenergy} the variance of $\bra{n}{\hat{S}_i^z}\ket{n}$ for eigenstates
$\ket{n}$ at virtually the same energy decreases with system size for a wide range of the spectrum. It should be
noted that the energy window in which the $\gtrsim 50$ eigenstates closest to the target energy lie
decreases with system size, clearly visible by inspecting for example the $L=12$ and $L=16$ data.
However, comparing data for adjacent targets suggests that the decreasing
variance\cite{ikeda_eigenstate_2011,khatami_quantum_2012,beugeling_finite-size_2014} of the local
magnetization is not an artefact of the decreasing energy window size but rather a generic feature,
leading to a well defined decrease of adjacent state local magnetization differences as shown in
Fig. \ref{fig:localmag} with a power law in the dimension of the Hilbert space, that we analyze in Fig. \ref{fig:diff_scaling} and discuss below.

At very high energy density, this is no longer true and the variance of this local observable over different
eigenstates remains very large, thus breaking ETH. This is fully consistent with the many body
mobility edge mapped in Ref. \onlinecite{luitz_many-body_2015}. Nevertheless, one should remain cautious
with this observation as for the present disorder strength $h=1.0$ the mobility edge lies close to
the boundary of the spectrum where finite size effects are expected to be large. Similarly, one
expects a mobility edge at the low end of the spectrum, which is even more difficult to resolve due
to stronger finite size effects, nevertheless the variance of $\bra{n}\hat{S}_5^z\ket{n}$ does not
seem to decrease between the two largest system sizes around $\epsilon\approx0.1$.

In order to study the decreasing variance of the local magnetization more rigorously, we calculate
the disorder average of the difference of $\hat{S}_i^z$ between an eigenstate $\ket{n}$ and the
adjacent eigenstate $\ket{n+1}$. This quantity is a measure for the variance of
the distribution of eigenstate expectation values of $\hat{S}_i^z$ over eigenstates within the
microcanonical energy window around the target energy and over disorder realizations. It is obvious
that the difference of local operators in adjacent eigenstates has to vanish if the quantity is to
become a smooth function of energy in the thermodynamic limit. In Fig. \ref{fig:localmag}, we show
the disorder averaged difference of local magnetizations of adjacent eigenstates as a function of
disorder strength for different system sizes. The signature of ETH is very clear as at weak disorder
strength these differences scale to zero with increasing system size, whereas the breakdown of ETH
in the MBL phase is signalled by a constant average difference for all system sizes. The position of
the critical point that can be estimated from the point at which the ETH is no longer valid depends
on the energy density and is fully consistent with a mobility edge\cite{luitz_many-body_2015,bera_many-body_2015,mondragon-shem_many-body_2015}.

Previously, it has been established that the variance of eigenstate to eigenstate fluctuations
of local operators in generic ETH systems decreases exponentially with system size as the square
root\cite{ikeda_eigenstate_2011,beugeling_finite-size_2014} of the dimension of the Hilbert space
$\text{dim}(\mathcal{H})$. We verify this exponential law in Fig. \ref{fig:diff_scaling} by plotting
the average difference of adjacent eigenstate magnetizations $|\Delta_m| = | \bra{n} S_i^z \ket{n} -
\bra{n+1} S_i^z \ket{n+1}|$ as a function of the dimension of the Hilbert space
$\text{dim}(\mathcal{H})$ on a log-log scale. Our results are approximately linear and the
corresponding exponent can be estimated by the (discrete) logarithmic derivative shown in the inset.
We find exponents which are close to $-\frac 1 2$ for the largest system sizes, in agreement with the
expectation of the square root behavior. The quality of the result is unfortunately not sufficient
to back speculations if the deviations from an exponent of $-\frac 1 2$ are mere finite size effects
or may have a deeper origin in the tails of the distributions to be discussed in Sec.
\ref{sec:rare}.

\section{Generic few body operators}

\begin{figure}[h]
      \centering
      \includegraphics{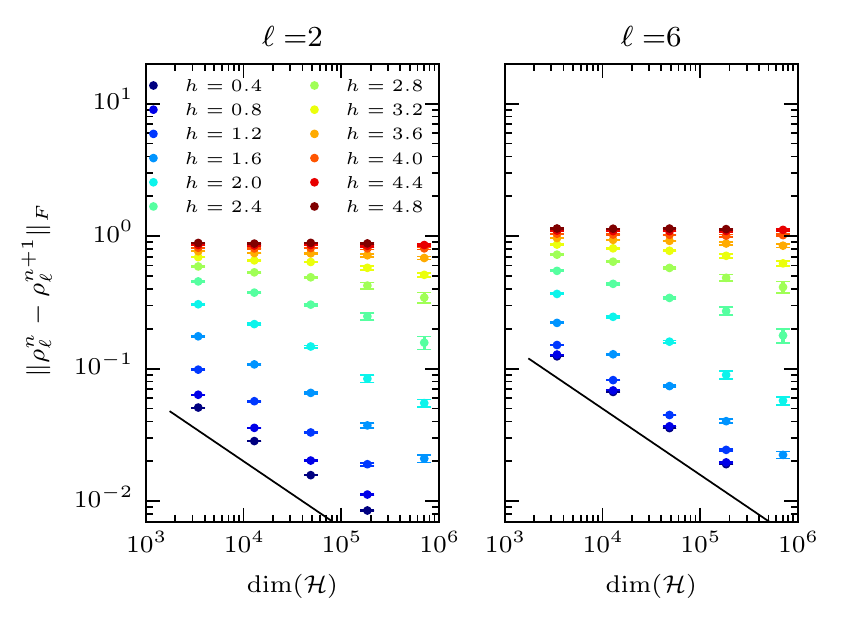}
      \caption{Disorder averaged norm of the difference of the reduced density matrices for adjacent
      eigenstates as a function of the dimension of the Hilbert space for two different subsystem
  sizes $\ell$ and different disorder strengths. The difference of the reduced density matrices
  decays as a power law in the dimension of the Hilbert space in the same way as the difference of
  local magnetizations in Fig. \ref{fig:diff_scaling}. The black line corresponds to
  $~\text{dim}(\mathcal{H})^{-\frac 1 2}$ for comparison and the exponent $-0.5$ is in good
  agreement with the results at weak disorder. In the MBL phase, the difference norm is constant
  with system size as expected.}
      \label{fig:generic_scaling}
  \end{figure}

  We have focussed on a specific local operator in the previous section: the local magnetization.
  This immediately leads to the question, whether different operators will also lead to a smooth
  function of energy and what condition these operators have to fulfill. In particular, we would
  like to investigate how ``local'' operators really have to be in order to fulfill ETH.
\begin{figure}[h]
    \centering
    \includegraphics{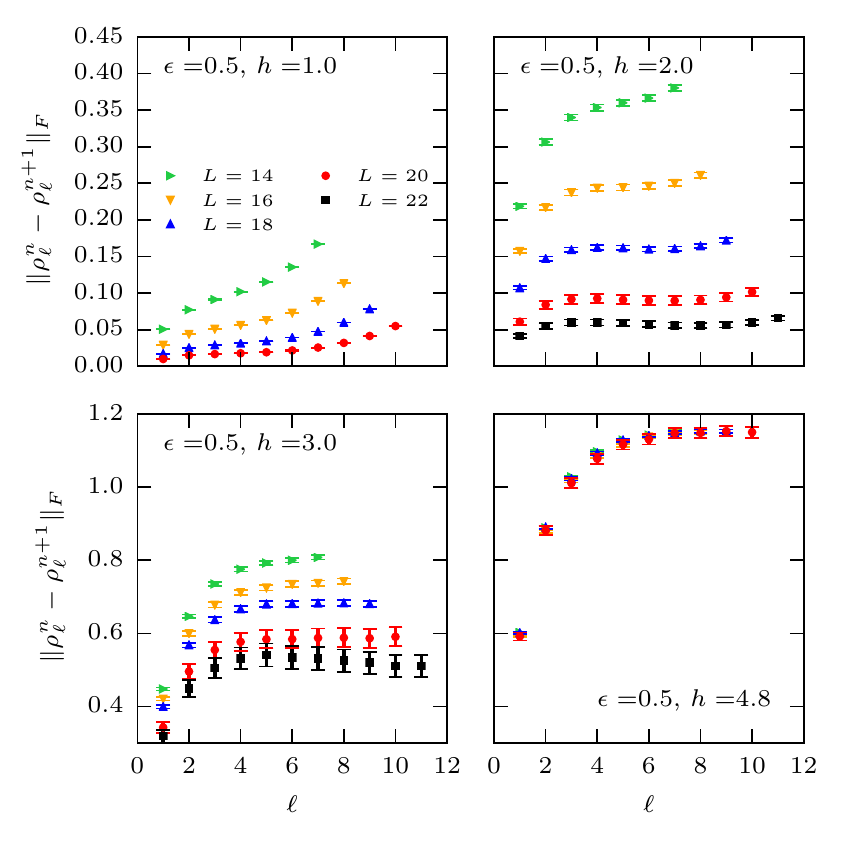}
    \caption{Disorder averaged Frobenius norm of the difference of reduced density matrices of a
    subsystem of size $\ell$ as a function of subsystem size.}
    \label{fig:generic}
\end{figure}
  All possible operators of a given support $\ell$ in two adjacent eigenstates $\ket{n}$ and
  $\ket{n+1}$ may be compared most efficiently by comparing the reduced density
  matrices\cite{alba_eigenstate_2015,garrison_does_2015}
  \begin{equation}
      \rho_\ell^n = \tr_{\overline{A_\ell}} \ket{n}\bra{n},
  \end{equation}
  where degrees of freedom corresponding to the \emph{complement} of the subsystem $A_\ell$ are
  traced out. The expectation values in state $\ket{n}$ of all operators involving only degrees of
  freedom on $A_\ell$ are fully determined by $\rho_\ell^n$ and therefore if ETH requires them to
  become a smooth function of energy, the corresponding reduced density matrices of adjacent
  eigenstates have to become \emph{equal} in the thermodynamic limit at least for ``small''
  subsystem sizes. In Fig. \ref{fig:generic}, we show the Frobenius norm $\| \rho_\ell^n -
  \rho_\ell^{n+1} \|_F = \sqrt{ \tr\left[( \rho_\ell^n - \rho_\ell^{n+1})^\dagger ( \rho_\ell^n -
  \rho_\ell^{n+1})\right]}$ of the difference of the reduced density matrices of adjacent eigenstates from
  the center of the spectrum ($\epsilon=0.5$), with the \emph{same} subsystem of size $\ell$ as a
  function of $\ell$ (up to $\ell=L/2$) and observe that the reduced density
  matrices become indeed \emph{equal} in the thermodynamic limit for astonishingly large subsystem
  sizes at weak disorder. This is quantitatively verified in Fig. \ref{fig:generic_scaling}, where
  we show the average reduced density matrix difference norm $\| \rho_\ell^n - \rho_\ell^{n+1} \|_F$ as a
  function of the dimension of the Hilbert space for different disorder strengths and $\ell=2$
  (left) together with $\ell=6$ (right). As for the fluctuations of the local magnetization, the
  difference norm scales to zero as a power law with an exponent close to $-\frac 1 2$ (the inverse
  square root law is indicated by the slope of the black line).
  For intermediate disorder strengths the absolute value of the difference norm is
  larger but still decreases with system size, in agreement with the expectation that closer to the
  critical point systems with a fixed size $L$ suffer from larger finite size effects due to a
  larger correlation length. 

  We conclude that expectation values of operators with a support smaller than half of the system
  size may lead to smooth functions of energy in the ETH phase. Close to a subsystem size of
  $\ell=L/2$, we observe an increase of the difference norm with subsystem size, however we find the
  difference norm to be still smaller than the $L/2$ difference norm of the previous size.

\section{Probability distributions}
\label{sec:rare}

Up to here, we have focussed on the behavior of disorder averages and concluded that in the ergodic
phase of the random field Heisenberg chain ETH is valid \emph{on average}. Let us now consider the
approach to the thermodynamic limit in more detail and discuss the full probability distributions
of the local magnetization difference $\Delta_{n,n+1} S_i^z$ of adjacent eigenstates as well as the
entanglement entropy, showing that rare states exist which do not obey ETH in the sense that their
fluctuations of local operators deviate from the normal
distribution\cite{beugeling_finite-size_2014} and that their
weight will become more important with growing disorder strength. It is interesting to note that the
presence of these non-ETH states coincides with the region of intermediate disorder of the phase
diagram in which subdiffusive transport is observed. At the same time, we find that the probability
distribution of the entanglement entropy exhibits low entanglement tails. The proposed mechanism for
the explanation of subdiffusion relies on the argument that rare Griffiths region of localized spins
exist, which act as bottlenecks for transport\cite{vosk_theory_2015,potter_universal_2015}. We
expect that these localized regions lead to lower entanglement entropies if the entanglement cut
lies within them and the observed low entanglement tails of the distribution are consistent with
this scenario.

\subsection{Probability distributions of eigenstate to eigenstate fluctuations}

\begin{figure}[h]
      \centering
      \includegraphics{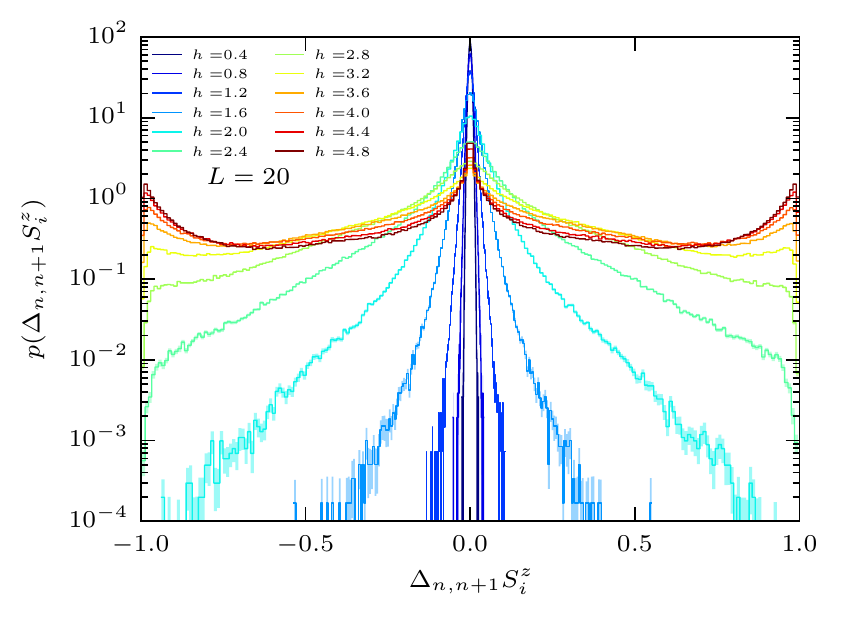}
      \caption{Probability density of the difference $\Delta_{n,n+1} S_i^z$ of the local magnetization $S_z^i$ of two
      eigenstates with adjacent energies for different disorder strenghts. The histograms are
  obtained from the central ($\epsilon=0.5$) 50 eigenstates of $10^3$ disorder realizations and
  contain the results for each site $i$ on which the local magnetization is compared.
  Even at very weak disorder, the distribution is already leptokurtic, showing heavier tails than
  the normal distribution. In the limit of very strong disorder, the distribution is dominated by
  the bimodal distribution of the local magnetization itself, which leads to the three dominant
  differences $-1, 0, 1$ stemming from random configurations of $S_z^i = -0.5, 0.5$ in the MBL
  phase.
  }
      \label{fig:mag_diff_histo}
  \end{figure}

\begin{figure}[h]
      \centering
      \includegraphics{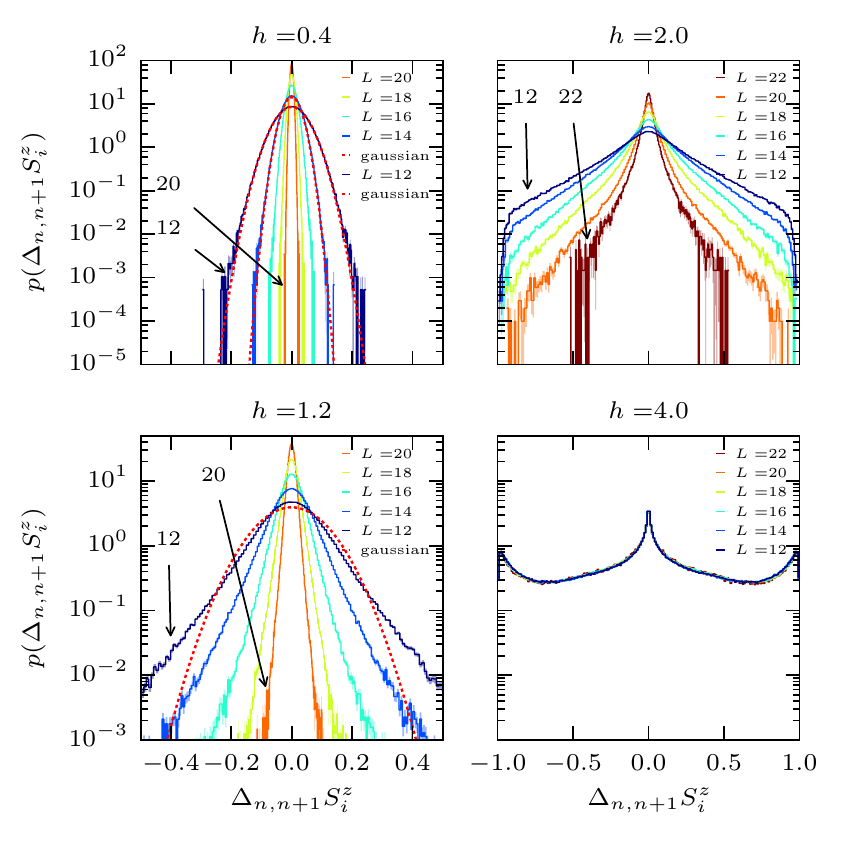}
      \caption{Probability density of the difference $\Delta_{n,n+1} S_i^z$ of the local
  magnetization $S_z^i$ of two eigenstates with adjacent energies for different system sizes and
  disorder strengths. For weak disorder, the reduction of the variance with system size is evident,
  however the distributions deviate increasingly from a gaussian distribution with growing disorder.
  In the MBL phase, the distributions become independent of system size, only governed by the
  localization length, which creates the nonzero weight between the three peaks of the distribution.}
      \label{fig:mag_diff_histo_size}
  \end{figure}

Let us first focus on the probability distribution of the differences of local magnetizations
$\Delta_{n,n+1} S_i^z = |\bra{n} S_i^z \ket{n} - \bra{n+1} S_i^z \ket{n+1}|$ of adjacent eigenstates
in the center of the spectrum ($\epsilon=0.5$) for different disorder strengths, shown in Fig.
\ref{fig:mag_diff_histo} for $L=20$. The argument in
Ref. \onlinecite{beugeling_finite-size_2014} that eigenstate expectation values of local operators
are effectively governed by the central limit theorem in generic ETH systems implies that one
expects that the differences of the local magnetization in adjacent eigenstates are distributed
according to a normal distribution with exponentially decreasing variance with system size and zero
mean, while the microscopic origin of the randomness that justifies the applicability of the central
limit theorem remains unclear.
This is in fact verified to high accuracy at very weak disorder strength. For example in the
top left panel of  Fig.  \ref{fig:mag_diff_histo_size} for $h=0.4$ we show the histogram of
$\Delta_{n,n+1} S_i^z$ for different system sizes and states in the center of the spectrum
($\epsilon=0.5$) together with best fits of the normal distribution, which seem to match perfectly
with the numerical data.

However, starting at relatively small disorder strengths, deviations from the normal distribution
emerge, as the distribution develops tails that are significantly heavier than those of the gaussian
distribution.

  \begin{figure}[h]
      \centering
      \includegraphics{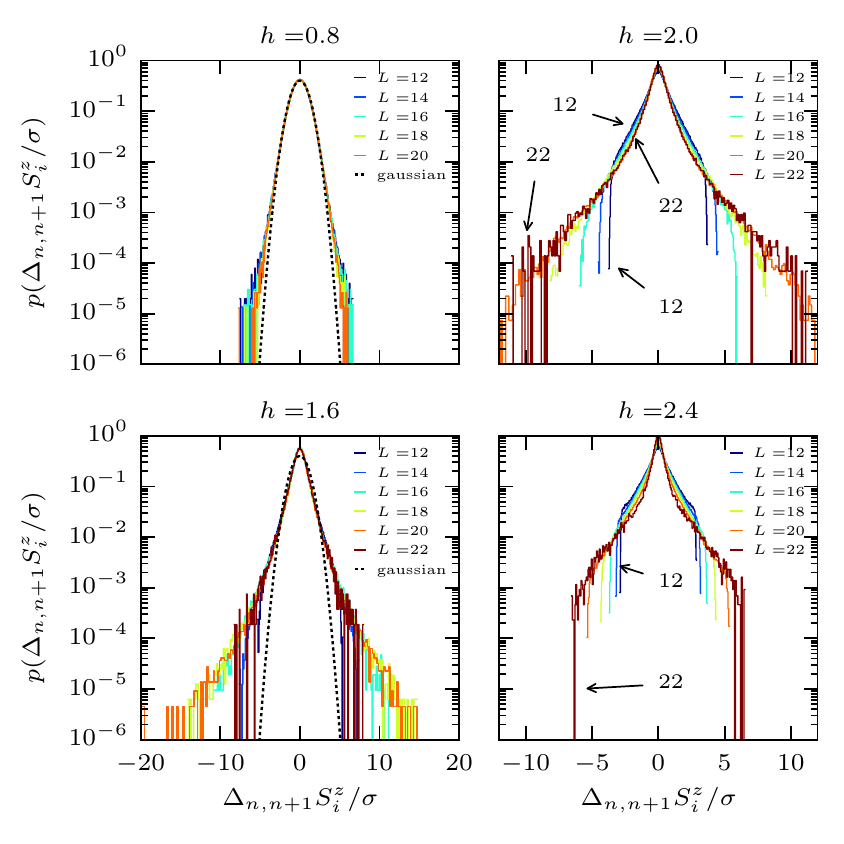}
      \caption{Probability density of the difference $\Delta_{n,n+1}S_i^z$ of the local
  magnetization $S_i^z$ of two eigenstates with adjacent energies, normalized by the standard
  deviation $\sigma$ to compare the shape of the distributions for different system sizes. For
  $h=0.8$, the weight of the tails seems to decrease with system size and approach the gaussian
  distribution. This is no longer true for stronger disorder and the weight of the tails increases
  strongly starting from $h\gtrsim 1.6$, accompanied by a growing kurtosis with system size
  in the same regime, making the distributions strongly leptokurtic. The kurtosis is
  defined by the fourth central moment divided by the squared variance, yielding a value of 3 for the gaussian
  distribution and larger values if the distribution has heavier tails, \ie if it is leptokurtic.}
      \label{fig:mag_diff_tail}
  \end{figure}

  While for weak disorder the deviations from the gaussian distribution seem to shrink with size, as seen \eg in Fig.
  \ref{fig:mag_diff_tail} for $h=0.8$, where we normalize the histograms by the standard deviation
  $\sigma$ to compare the shapes, this is no longer true at intermediate disorder, where the
  deviations from the gaussian distribution increase.
  In fact, for disorder strengths $h\gtrsim1.6$, the weight of the
  tails actually grows with system size, pointing clearly to an effect that survives in the
  thermodynamic limit. This is most prominent in Fig. \ref{fig:mag_diff_tail} for $h=2.0$ and
  $h=2.4$, where it seems that for the largest sizes we divide by a \emph{too large standard
  deviation}, thus deforming the shape of the central peak, which narrows with system size, while the
  tails increase. This means that the variance is in fact larger than one would expect and is
  therefore influenced by the tails of the distribution, hinting for an effect that survives in the
  thermodynamic limit. On a more speculative note, this could be related to a slower scaling of the
  variances of the distribution to the thermodynamic limit as discussed in Figs.
  \ref{fig:diff_scaling} and \ref{fig:generic_scaling}, modifying the exponent.

  Interestingly, the distribution in the MBL phase at strong disorder becomes trimodal, indicating
  that local magnetizations of adjacent eigenstates are close to $\bra{n} S_i^z \ket{n} = \pm \frac
  1 2$ and \emph{completely independent}, thus suggesting a binomial distribution at strong disorder
  \footnote{This is more apparent for larger subsystem magnetizations, where the distribution has a
      larger number of peaks.}, dressed by slight deviations caused by the finite support of the l-bits in terms of the
  localization length. As the localization length seems to be smaller than our system sizes already
  at $h=4.0$, we observe a distribution virtually independent of $L$ in Fig.
  \ref{fig:mag_diff_histo_size}.

\subsection{Probability distributions of the entanglement entropy}

  One of the most studied quantities for the MBL transition is certainly the entanglement entropy,
  which is known to change its mean behavior from a thermal volume law in the ETH phase to an area law in
  the MBL phase. However, the probability distributions of the entanglement entropy have received
  less attention until recently\cite{vosk_theory_2015,baygan_many-body_2015-1,lim_nature_2015}, possibly due to the daunting requirement for
  the number of disorder realizations to obtain a satisfactory resolution of the histogram. Here, we
  show results for the probability distribution of the von Neumann entanglement entropy
  $S_\text{E}=-\tr_A\left[ (\tr_B \ket{n}\bra{n}) \ln (\tr_B \ket{n}\bra{n})\right]$, where the
  subsystem $A$ represents half of the chain and $B$ is the complement of $A$. Averaging over
  approximately $>10^3$ disorder realizations, $~50$ eigenstates per realization from the center of
  the spectrum ($\epsilon=0.5$) and over \emph{all cuts} with subsystem size $L/2$, we can generate
  histograms from roughly $10^6$ samples of the entanglement entropy.

\begin{figure}[h]
    \centering
    \includegraphics{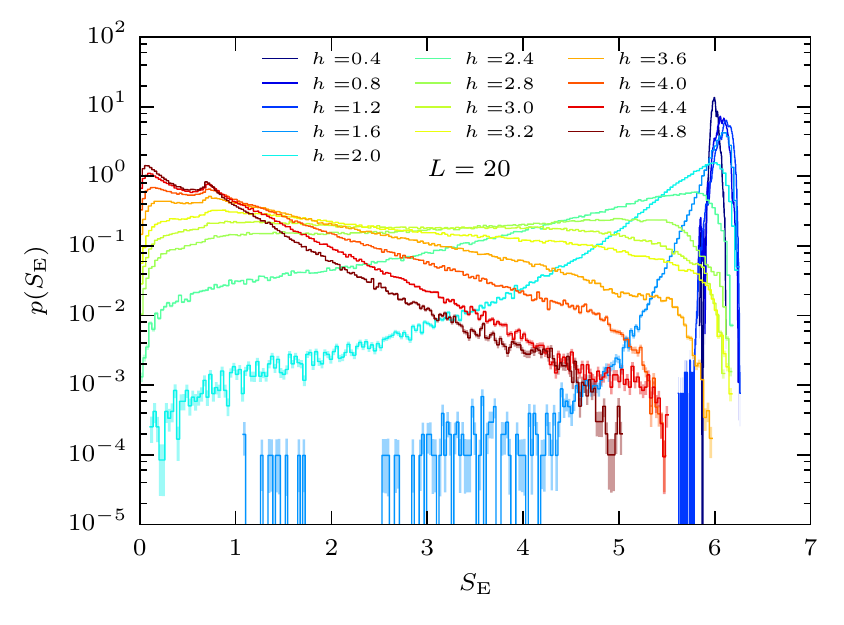}
    \caption{ Probability density of the entanglement entropy $S_E$ of states in the middle of the
        spectrum ($\epsilon=0.5$) for different disorder strengths. We use the bipartition $L_A=L_B$
        for the periodic $L=20$ chain. The evolution from the volume law behavior at weak disorder
        to the area law at strong disorder is clearly visible. At very weak disorder, the
        distribution is close to gaussian and develops exponential tails at intermediate disorder.
        They correspond to rare regions with low entanglement, responsible for slow (subdiffusive)
        transport.
        }
    \label{fig:ee_histo}
\end{figure}
    The resulting histograms for different disorder strengths are shown in Fig.
    \ref{fig:ee_histo}. For very weak disorder, the distribution is close to a normal distribution
    and the mode (position of the maximum) of the distribution increases slightly with increasing disorder, possibly due to
    the proximity to the integrable point at $h=0$, where additional integrals of motion restrict
    the degrees of freedom of eigenstates in the Hilbert space.

    At weak to intermediate disorder strengths, the mode of the
    distribution is virtually unchanged but a long tail develops, reaching to extremely low values
    of entanglement. This corresponds to positions of the subsystem cut, where the two subsystems
    are virtually decoupled, leading to a product state of two smaller system eigenstates and a weak
    link between them. The weight of these very low entanglement cuts increases dramatically when
    approaching the transition, leading to a broad distribution with maximal variance around $h=3$.
    This distribution corresponds to the observed maximal variance of the entanglement entropy in
    the proximity of the critical point\cite{kjall_many-body_2014,luitz_many-body_2015,chen_many-body_2015}.

    In the MBL phase, the mode of the distribution shifts to a very low (constant) value, creating
    the area (constant) law of the mean and develops a resonance at $S_\text{E} \approx \ln 2$,
    before showing a long exponential decay with arbitrarily strong entanglement. The peak close to
    zero entanglement reflects the fact that states are effectively product states in the l-bit
    basis, which becomes identical with the real space basis if the localization length becomes
    small enough at large disorder.

    These results suggest that the
    environment of the transition in both ETH and MBL phases is governed by rare event effects,
    although their importance in the MBL phase is not clear as they may not be dominating the
    overall physics. The regime close to the critical point, dominated by Griffiths effects, has also
    recently been studied using a matrix product state based method to access larger system sizes in
    Ref. \onlinecite{lim_nature_2015} together with exact diagonalization and the exponential tail
    of the distribution was also observed as well as a broad histogram\footnote{It should
    be noted that the differences in the details of the distributions in
    Ref. \onlinecite{lim_nature_2015} and the present work are due to different boundary conditions: Here
    we use periodic boundaries, while Ref. \onlinecite{lim_nature_2015} employ open boundary
    conditions.} at $h=3$.

\begin{figure}[h]
    \centering
    \includegraphics{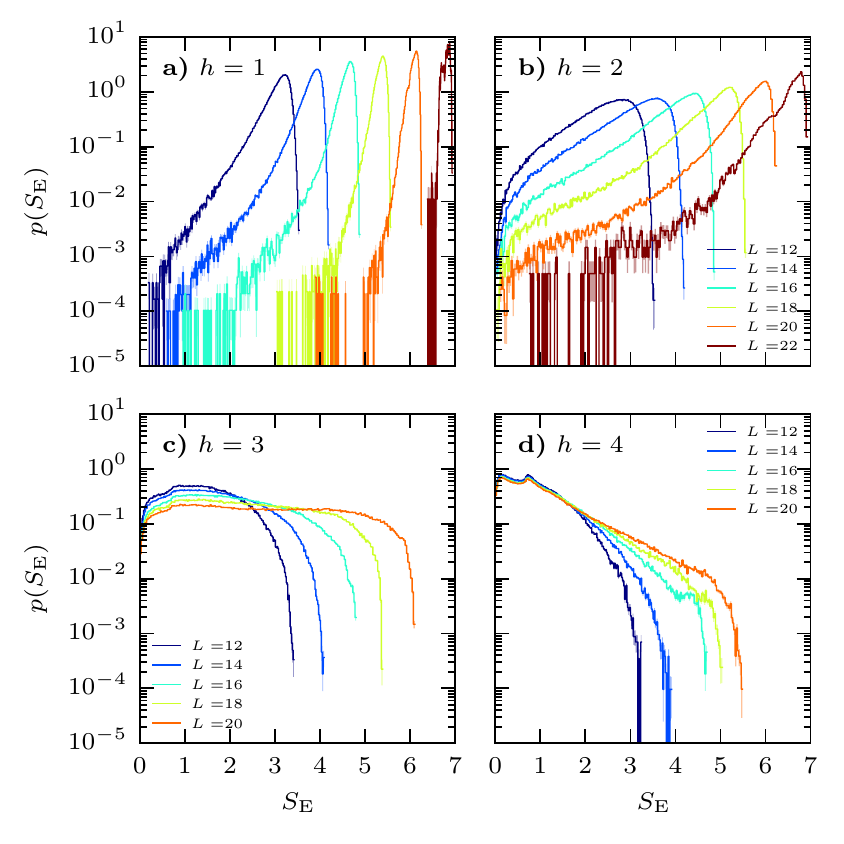}
    \caption{Probability density of the entanglement entropy $S_E$ of states in the middle of the
    spectrum for different system sizes at disorder strenghts a) $h=1$, b) $h=2$, c) $h=3$ and
d) $h=4$.}
    \label{fig:ee_histo_L}
\end{figure}

    Here, we try to connect the presence of low entanglement entropies, to the observed subdiffusive
    transport regime\cite{lim_nature_2015,vosk_theory_2015,potter_universal_2015,agarwal_anomalous_2015,luitz_extended_2016}
    at intermediate disorder. The proposed mechanism for subdiffusion is via rare nearly localized
    regions, acting as bottlenecks for transport and the entanglement growth in time. If such
    regions exist, one would naturally expect that the entanglement entropy of the corresponding
    eigenstates will be low if the cut between the subsystems lies in such a rare region.
    In fact, one should note that one single localized region will only reduce the entanglement entropy by less
    than a factor of two, as it reduces entanglement only over one boundary of the subsystem down to
    the typical constant MBL entanglement entropy. As the total system has periodic boundaries, the
    second half of the entanglement entropy can still stem from the other subsystem boundary.
    Together with this argument, we conclude that we find the expected low entanglement entropies in the
    regime in which transport is subdiffusive and that our results are consistent with this
    scenario. In the next section, we will discuss this in further detail.

Let us briefly discuss the
dependence on system size of the distributions of the entanglement entropy, shown in Fig.
\ref{fig:ee_histo_L}. At weak disorder strength ($h=1.0$), we observe a clear volume law and the
importance of the low entanglement tail seems to decrease with growing system size, approaching more
and more a normal distribution, although all the
data for $L=22$ should be treated with care due to a number of samples that is by an order of
magnitude smaller than for the other system sizes. At $h=2$ it is clear that the tail of
the distribution extends all the way down to very low entanglement. On some very rare occasions, we
observe nearly zero entanglement, which can only occur if the eigenstate is close to a product state
or in other words if \emph{two} localized regions fall exactly on the cut between the subsystems.

Interestingly, at $h=3$, before the estimated location of the critical point at $h\approx3.7$, the
distribution seems to become very broad, bounded by the maximal entanglement entropy, 
which is the reason for the observed maximum of the variance of entanglement entropy close to the critical
point\cite{kjall_many-body_2014,luitz_many-body_2015,chen_many-body_2015}, here the weight of low
entanglement states becomes very large and the behavior is dominated by rare region effects as
discussed in Ref. \onlinecite{lim_nature_2015}. The dependence on system size suggests, however,
that for even larger systems the mode of the distribution can still shift to the thermal value,
leaving a large weight at low entanglement entropies.

In the MBL phase ($h=4$), the mode of the distribution is at a low (constant with system size) value, however the tail of the
distribution grows with system size, reaching possibly up to the maximal entanglement entropy given by $L/2 \ln 2$.

\subsection{Weak links}

Let us finally consider the spatial entanglement structure of the central eigenstates
($\epsilon=0.5$) in a typical disorder realization of length $L=22$ at intermediate disorder
strength $h=2.0$.

\begin{figure}[h]
    \centering
    \includegraphics{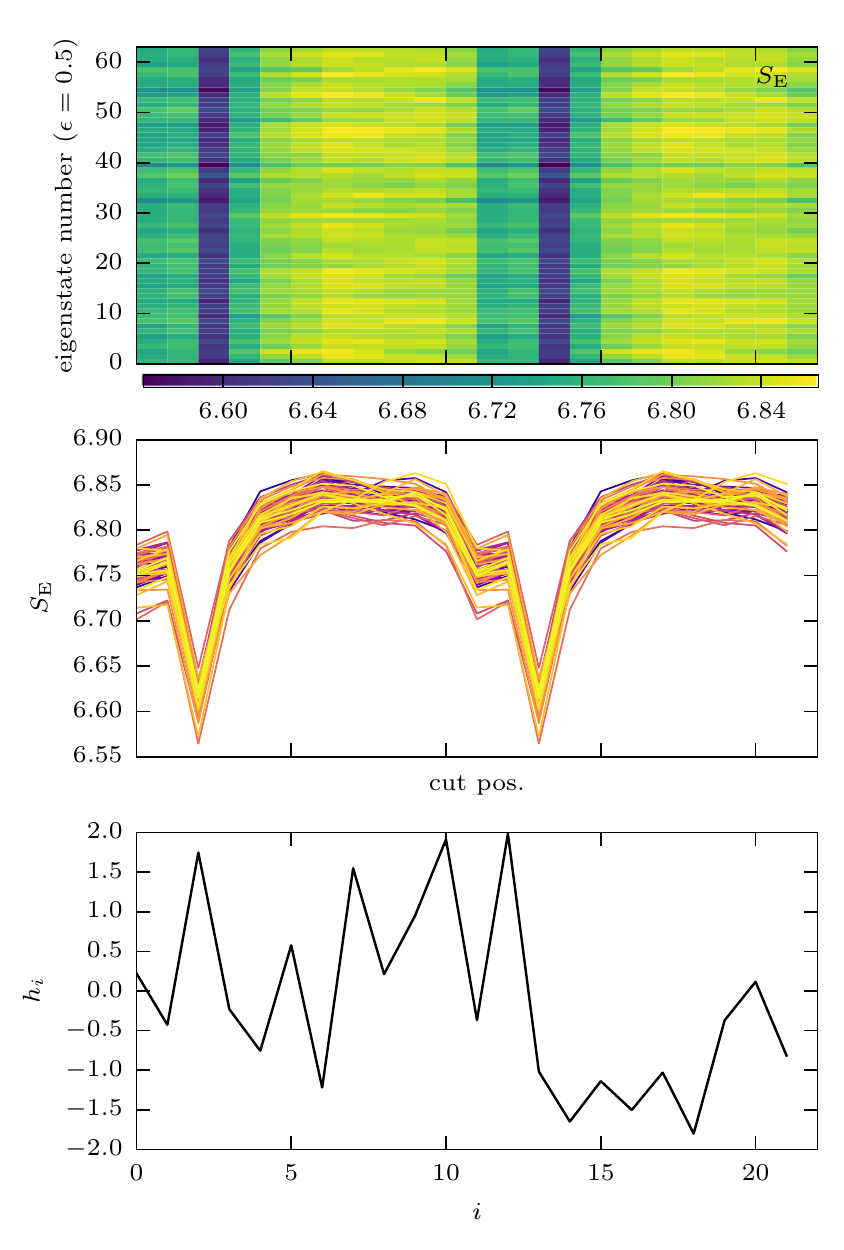}
    \caption{Top panel: Spatial variation of the entanglement entropy $S_E$ for the central 63 eigenstates
        of one disorder realization $h=2$ in the periodic $L=22$ chain (The disorder configuration ${h_i}$ as a
    function of the site index $i$ is shown in the bottom panel.) as a function of the position of
the cut between subsystem $A$ and $B$. The subsystem size is $\ell=L/2$. Center panel: Same data as
in the top panel as line graph. Depending on the cut position, the entanglement entropy varies and
is found to be particularly low for some cuts, corresponding to a weaker link between the
subsystems.  }
    \label{fig:entanglement_profile}
\end{figure}

We show in the top panel of Fig. \ref{fig:entanglement_profile} for the central 63 eigenstates the
entanglement entropy as a function of the cut position for all 22 possible cut positions (of which
the second half is naturally equivalent to the first half). This representation reveals an
intriguing feature: all states show a minimal entanglement entropy if the system is cut between 
spins 1 and 2 (the second cut is between spins 12 and 13), while the
entanglement entropy is larger for other cut positions. Although in this sample the drop in the
entanglement entropy is not very large, it is interesting to observe that the spatial structure is
\emph{identical} in all eigenstates in the energy window. This means that the entanglement across one of the
boundaries of the subsystem is slightly smaller for all eigenstates and could be caused by a weak
link, possibly leading to slightly slower transport. One would expect that such a weak link is caused by a field
configuration that localizes spins through large fields and the corresponding configuration in the
lower panel of Fig. \ref{fig:entanglement_profile} suggests that this may be caused by the large
field $h_i$ on spin 12 and to a lesser extent on spin 2, in fact, a comparison of many samples leads
to the hypothesis that a large change in the field configuration may lead to a weak link.

\begin{figure}[h]
    \centering
    \includegraphics{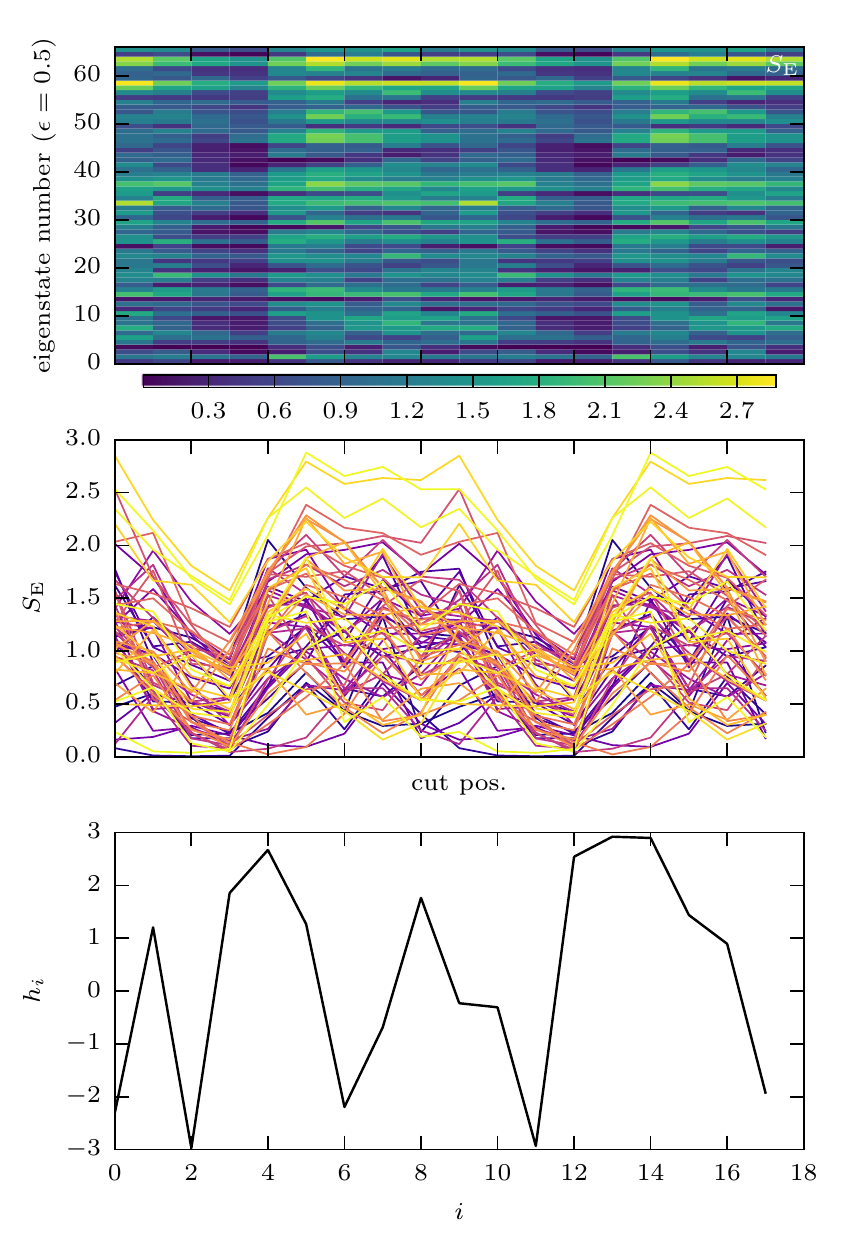}
    \caption{Top panel: Spatial variation of the entanglement entropy $S_E$ for the central 67 eigenstates
        of one disorder realization $h=3$ in the periodic $L=18$ chain (The disorder configuration ${h_i}$ as a
    function of the site index $i$ is shown in the bottom panel.) as a function of the position of
the cut between subsystem $A$ and $B$. The subsystem size is $\ell=L/2$. Center panel: Same data as
in the top panel as line graph. Depending on the cut position, the entanglement entropy varies and
is found to be particularly low for some cuts, corresponding to a weaker link between the
subsystems. }
    \label{fig:entanglement_profile2}
\end{figure}

In Fig. \ref{fig:entanglement_profile2}, we present a sample at $h=3$, $L=18$, where the effect is
much more severe: Here, the entanglement entropy is reduced roughly by a factor of 2 for most states
if the cut lies between spins 2 and 3 and spins 11 and 12, compared to other cuts. Again, the
spatial correlation between all states is visible and maximal field changes on both boundaries of the
subsystem at spins 2 and 11 may be responsible for the weak link.

Let us also mention that the correlation of the entanglement profile in all eigenstates has also
been observed in Ref. \onlinecite{bera_local_2015} by the nearest neighbor concurrence as a measure of
entanglement. Our results show that the low entanglement tails in the distribution of the
entanglement entropy in the ETH phase are not simply caused by low entanglement states (which do
also exist) but are at least partly created by a spatial variation of the entanglement entropy as a
function of the positions of the subsystem boundaries. The effect is the strongest if at least two localized
regions are present at a distance that corresponds to the subsystem length.

\section{Conclusion}

  We have studied in detail the validity of the eigenstate thermalization hypothesis in the random
  field Heisenberg chain and find that while typical states in the ergodic phase obey ETH, there are
  rare states with local expectation values far from the ETH mean, 
  leading to tails of the fluctuation distributions that are significantly heavier than those of the
  normal distribution.
 
  A comparison of distributions for different system sizes shows that the ETH violating tails can
  survive in the thermodynamic limit, as at intermediate disorder strength the weight of the tails
  can even increase with system size. This coincides with the region of the phase diagram that
  has been reported to be dominated by Griffiths
  effects\cite{agarwal_anomalous_2015,vosk_theory_2015,potter_universal_2015,luitz_extended_2016,laumann_many-body_2014,lim_nature_2015}
  close to the MBL transition.

  Similar rare event tails are also observed in the distribution of the entanglement entropy, which
  become dominant close to the critical point, creating the large variance of the entanglement
  entropy that has been reported
  previously\cite{kjall_many-body_2014,luitz_many-body_2015,chen_many-body_2015}. We argue that at
  least a part of the weight of the tails of the distribution is caused by the spatial entanglement
  structure, showing typical disorder realizations that have a large intrinsic variance of the
  entanglement entropy, when changing the position of the cut between the subsystems. This
  observation is in agreement with the proposed mechanism for subdiffusive transport, which relies
  on rare localized regions that act as weak links.
  
  While it is consistent with other studies that low entanglement regions in the ergodic phase may
  be the cause of the anomalous transport properties at intermediate disorder, it remains an open
  question if the MBL phase has pathological properties close to the transition due to rare high
  entanglement regions. The presently accessible system sizes also leave the question of the shape
  of the distribution at the critical point unanswered. 
  On a final note, we mention that close to the transition, the observed high
  entanglement tails in the MBL phase might be problematic for matrix product state based
  methods\cite{yu_finding_2015,khemani_obtaining_2015,lim_nature_2015,kennes_entanglement_2015},
  requiring larger bond dimensions.

\begin{acknowledgments}
    It is a great pleasure to thank Fabien Alet and Nicolas Laflorencie for many fruitful
    collaborations and their critical comments on the manuscript. 
    The author is grateful to Jens Bardarson, Bryan Clark, Hitesh Changlani, Eduardo Fradkin, Markus
    Heyl, Cecile Monthus, Lode Pollet and Frank Pollmann for helpful discussions and to Luiz Santos
    for his critical reading of the article as well as for many discussions.
    This work was supported in part by the Gordon and Betty Moore Foundation's EPiQS Initiative
    through Grant No. GBMF4305 at the University of Illinois and the French ANR program
    ANR-11-IS04-005-01.
    The code is based on the PETSc~\cite{petsc-web-page,petsc-user-ref,petsc-efficient}, 
    SLEPc~\cite{hernandez_slepc:_2005} and MUMPS\cite{MUMPS1,MUMPS2} libraries and calculations were performed using HPC
    resources from CALMIP (grant 2015-P0677).
    This research is part of the Blue Waters sustained-petascale computing project, which is
    supported by the National Science Foundation (awards OCI-0725070 and ACI-1238993) and the state
    of Illinois. Blue Waters is a joint effort of the University of Illinois at Urbana-Champaign and
    its National Center for Supercomputing Applications.
\end{acknowledgments}

\bibliography{eth,mbl_eth}

\begin{thebibliography}{57}%
\makeatletter
\providecommand \@ifxundefined [1]{%
 \@ifx{#1\undefined}
}%
\providecommand \@ifnum [1]{%
 \ifnum #1\expandafter \@firstoftwo
 \else \expandafter \@secondoftwo
 \fi
}%
\providecommand \@ifx [1]{%
 \ifx #1\expandafter \@firstoftwo
 \else \expandafter \@secondoftwo
 \fi
}%
\providecommand \natexlab [1]{#1}%
\providecommand \enquote  [1]{``#1''}%
\providecommand \bibnamefont  [1]{#1}%
\providecommand \bibfnamefont [1]{#1}%
\providecommand \citenamefont [1]{#1}%
\providecommand \href@noop [0]{\@secondoftwo}%
\providecommand \href [0]{\begingroup \@sanitize@url \@href}%
\providecommand \@href[1]{\@@startlink{#1}\@@href}%
\providecommand \@@href[1]{\endgroup#1\@@endlink}%
\providecommand \@sanitize@url [0]{\catcode `\\12\catcode `\$12\catcode
  `\&12\catcode `\#12\catcode `\^12\catcode `\_12\catcode `\%12\relax}%
\providecommand \@@startlink[1]{}%
\providecommand \@@endlink[0]{}%
\providecommand \url  [0]{\begingroup\@sanitize@url \@url }%
\providecommand \@url [1]{\endgroup\@href {#1}{\urlprefix }}%
\providecommand \urlprefix  [0]{URL }%
\providecommand \Eprint [0]{\href }%
\providecommand \doibase [0]{http://dx.doi.org/}%
\providecommand \selectlanguage [0]{\@gobble}%
\providecommand \bibinfo  [0]{\@secondoftwo}%
\providecommand \bibfield  [0]{\@secondoftwo}%
\providecommand \translation [1]{[#1]}%
\providecommand \BibitemOpen [0]{}%
\providecommand \bibitemStop [0]{}%
\providecommand \bibitemNoStop [0]{.\EOS\space}%
\providecommand \EOS [0]{\spacefactor3000\relax}%
\providecommand \BibitemShut  [1]{\csname bibitem#1\endcsname}%
\let\auto@bib@innerbib\@empty
\bibitem [{\citenamefont {Deutsch}(1991)}]{deutsch_quantum_1991}%
  \BibitemOpen
  \bibfield  {author} {\bibinfo {author} {\bibfnamefont {J.~M.}\ \bibnamefont
  {Deutsch}},\ }\bibfield  {title} {\enquote {\bibinfo {title} {Quantum
  statistical mechanics in a closed system},}\ }\href {\doibase
  10.1103/PhysRevA.43.2046} {\bibfield  {journal} {\bibinfo  {journal} {Phys.
  Rev. A}\ }\textbf {\bibinfo {volume} {43}},\ \bibinfo {pages} {2046--2049}
  (\bibinfo {year} {1991})}\BibitemShut {NoStop}%
\bibitem [{\citenamefont {Srednicki}(1994)}]{srednicki_chaos_1994}%
  \BibitemOpen
  \bibfield  {author} {\bibinfo {author} {\bibfnamefont {Mark}\ \bibnamefont
  {Srednicki}},\ }\bibfield  {title} {\enquote {\bibinfo {title} {Chaos and
  quantum thermalization},}\ }\href {\doibase 10.1103/PhysRevE.50.888}
  {\bibfield  {journal} {\bibinfo  {journal} {Phys. Rev. E}\ }\textbf {\bibinfo
  {volume} {50}},\ \bibinfo {pages} {888--901} (\bibinfo {year}
  {1994})}\BibitemShut {NoStop}%
\bibitem [{\citenamefont {D'Alessio}\ \emph {et~al.}(2015)\citenamefont
  {D'Alessio}, \citenamefont {Kafri}, \citenamefont {Polkovnikov},\ and\
  \citenamefont {Rigol}}]{dalessio_quantum_2015}%
  \BibitemOpen
  \bibfield  {author} {\bibinfo {author} {\bibfnamefont {Luca}\ \bibnamefont
  {D'Alessio}}, \bibinfo {author} {\bibfnamefont {Yariv}\ \bibnamefont
  {Kafri}}, \bibinfo {author} {\bibfnamefont {Anatoli}\ \bibnamefont
  {Polkovnikov}}, \ and\ \bibinfo {author} {\bibfnamefont {Marcos}\
  \bibnamefont {Rigol}},\ }\bibfield  {title} {\enquote {\bibinfo {title} {From
  {Quantum} {Chaos} and {Eigenstate} {Thermalization} to {Statistical}
  {Mechanics} and {Thermodynamics}},}\ }\href {http://arxiv.org/abs/1509.06411}
  {\bibfield  {journal} {\bibinfo  {journal} {arXiv:1509.06411 [cond-mat,
  physics:quant-ph]}\ } (\bibinfo {year} {2015})},\ \bibinfo {note} {arXiv:
  1509.06411}\BibitemShut {NoStop}%
\bibitem [{\citenamefont {Rigol}\ \emph {et~al.}(2008)\citenamefont {Rigol},
  \citenamefont {Dunjko},\ and\ \citenamefont
  {Olshanii}}]{rigol_thermalization_2008}%
  \BibitemOpen
  \bibfield  {author} {\bibinfo {author} {\bibfnamefont {Marcos}\ \bibnamefont
  {Rigol}}, \bibinfo {author} {\bibfnamefont {Vanja}\ \bibnamefont {Dunjko}}, \
  and\ \bibinfo {author} {\bibfnamefont {Maxim}\ \bibnamefont {Olshanii}},\
  }\bibfield  {title} {\enquote {\bibinfo {title} {Thermalization and its
  mechanism for generic isolated quantum systems},}\ }\href {\doibase
  10.1038/nature06838} {\bibfield  {journal} {\bibinfo  {journal} {Nature}\
  }\textbf {\bibinfo {volume} {452}},\ \bibinfo {pages} {854--858} (\bibinfo
  {year} {2008})}\BibitemShut {NoStop}%
\bibitem [{\citenamefont {Rigol}(2009)}]{rigol_breakdown_2009}%
  \BibitemOpen
  \bibfield  {author} {\bibinfo {author} {\bibfnamefont {Marcos}\ \bibnamefont
  {Rigol}},\ }\bibfield  {title} {\enquote {\bibinfo {title} {Breakdown of
  {Thermalization} in {Finite} {One}-{Dimensional} {Systems}},}\ }\href
  {\doibase 10.1103/PhysRevLett.103.100403} {\bibfield  {journal} {\bibinfo
  {journal} {Phys. Rev. Lett.}\ }\textbf {\bibinfo {volume} {103}},\ \bibinfo
  {pages} {100403} (\bibinfo {year} {2009})}\BibitemShut {NoStop}%
\bibitem [{\citenamefont {Beugeling}\ \emph {et~al.}(2014)\citenamefont
  {Beugeling}, \citenamefont {Moessner},\ and\ \citenamefont
  {Haque}}]{beugeling_finite-size_2014}%
  \BibitemOpen
  \bibfield  {author} {\bibinfo {author} {\bibfnamefont {W.}~\bibnamefont
  {Beugeling}}, \bibinfo {author} {\bibfnamefont {R.}~\bibnamefont {Moessner}},
  \ and\ \bibinfo {author} {\bibfnamefont {Masudul}\ \bibnamefont {Haque}},\
  }\bibfield  {title} {\enquote {\bibinfo {title} {Finite-size scaling of
  eigenstate thermalization},}\ }\href {\doibase 10.1103/PhysRevE.89.042112}
  {\bibfield  {journal} {\bibinfo  {journal} {Phys. Rev. E}\ }\textbf {\bibinfo
  {volume} {89}},\ \bibinfo {pages} {042112} (\bibinfo {year}
  {2014})}\BibitemShut {NoStop}%
\bibitem [{\citenamefont {Steinigeweg}\ \emph {et~al.}(2013)\citenamefont
  {Steinigeweg}, \citenamefont {Herbrych},\ and\ \citenamefont {Prelov{\v
  s}ek}}]{steinigeweg_eigenstate_2013}%
  \BibitemOpen
  \bibfield  {author} {\bibinfo {author} {\bibfnamefont {R.}~\bibnamefont
  {Steinigeweg}}, \bibinfo {author} {\bibfnamefont {J.}~\bibnamefont
  {Herbrych}}, \ and\ \bibinfo {author} {\bibfnamefont {P.}~\bibnamefont
  {Prelov{\v s}ek}},\ }\bibfield  {title} {\enquote {\bibinfo {title}
  {Eigenstate thermalization within isolated spin-chain systems},}\ }\href
  {\doibase 10.1103/PhysRevE.87.012118} {\bibfield  {journal} {\bibinfo
  {journal} {Phys. Rev. E}\ }\textbf {\bibinfo {volume} {87}},\ \bibinfo
  {pages} {012118} (\bibinfo {year} {2013})}\BibitemShut {NoStop}%
\bibitem [{\citenamefont {Alba}(2015)}]{alba_eigenstate_2015}%
  \BibitemOpen
  \bibfield  {author} {\bibinfo {author} {\bibfnamefont {Vincenzo}\
  \bibnamefont {Alba}},\ }\bibfield  {title} {\enquote {\bibinfo {title}
  {Eigenstate thermalization hypothesis and integrability in quantum spin
  chains},}\ }\href {\doibase 10.1103/PhysRevB.91.155123} {\bibfield  {journal}
  {\bibinfo  {journal} {Phys. Rev. B}\ }\textbf {\bibinfo {volume} {91}},\
  \bibinfo {pages} {155123} (\bibinfo {year} {2015})}\BibitemShut {NoStop}%
\bibitem [{\citenamefont {Ikeda}\ \emph {et~al.}(2011)\citenamefont {Ikeda},
  \citenamefont {Watanabe},\ and\ \citenamefont
  {Ueda}}]{ikeda_eigenstate_2011}%
  \BibitemOpen
  \bibfield  {author} {\bibinfo {author} {\bibfnamefont {Tatsuhiko~N.}\
  \bibnamefont {Ikeda}}, \bibinfo {author} {\bibfnamefont {Yu}~\bibnamefont
  {Watanabe}}, \ and\ \bibinfo {author} {\bibfnamefont {Masahito}\ \bibnamefont
  {Ueda}},\ }\bibfield  {title} {\enquote {\bibinfo {title} {Eigenstate
  randomization hypothesis: {Why} does the long-time average equal the
  microcanonical average?}}\ }\href {\doibase 10.1103/PhysRevE.84.021130}
  {\bibfield  {journal} {\bibinfo  {journal} {Phys. Rev. E}\ }\textbf {\bibinfo
  {volume} {84}},\ \bibinfo {pages} {021130} (\bibinfo {year}
  {2011})}\BibitemShut {NoStop}%
\bibitem [{\citenamefont {Rigol}\ and\ \citenamefont
  {Srednicki}(2012)}]{rigol_alternatives_2012}%
  \BibitemOpen
  \bibfield  {author} {\bibinfo {author} {\bibfnamefont {Marcos}\ \bibnamefont
  {Rigol}}\ and\ \bibinfo {author} {\bibfnamefont {Mark}\ \bibnamefont
  {Srednicki}},\ }\bibfield  {title} {\enquote {\bibinfo {title} {Alternatives
  to {Eigenstate} {Thermalization}},}\ }\href {\doibase
  10.1103/PhysRevLett.108.110601} {\bibfield  {journal} {\bibinfo  {journal}
  {Physical Review Letters}\ }\textbf {\bibinfo {volume} {108}} (\bibinfo
  {year} {2012}),\ 10.1103/PhysRevLett.108.110601},\ \bibinfo {note} {arXiv:
  1108.0928}\BibitemShut {NoStop}%
\bibitem [{\citenamefont {Ikeda}\ and\ \citenamefont
  {Ueda}(2015)}]{ikeda_how_2015}%
  \BibitemOpen
  \bibfield  {author} {\bibinfo {author} {\bibfnamefont {Tatsuhiko~N.}\
  \bibnamefont {Ikeda}}\ and\ \bibinfo {author} {\bibfnamefont {Masahito}\
  \bibnamefont {Ueda}},\ }\bibfield  {title} {\enquote {\bibinfo {title} {How
  accurately can the microcanonical ensemble describe small isolated quantum
  systems?}}\ }\href {\doibase 10.1103/PhysRevE.92.020102} {\bibfield
  {journal} {\bibinfo  {journal} {Phys. Rev. E}\ }\textbf {\bibinfo {volume}
  {92}},\ \bibinfo {pages} {020102} (\bibinfo {year} {2015})}\BibitemShut
  {NoStop}%
\bibitem [{\citenamefont {Luitz}\ \emph {et~al.}(2015)\citenamefont {Luitz},
  \citenamefont {Laflorencie},\ and\ \citenamefont
  {Alet}}]{luitz_many-body_2015}%
  \BibitemOpen
  \bibfield  {author} {\bibinfo {author} {\bibfnamefont {David~J.}\
  \bibnamefont {Luitz}}, \bibinfo {author} {\bibfnamefont {Nicolas}\
  \bibnamefont {Laflorencie}}, \ and\ \bibinfo {author} {\bibfnamefont
  {Fabien}\ \bibnamefont {Alet}},\ }\bibfield  {title} {\enquote {\bibinfo
  {title} {Many-body localization edge in the random-field {Heisenberg}
  chain},}\ }\href {\doibase 10.1103/PhysRevB.91.081103} {\bibfield  {journal}
  {\bibinfo  {journal} {Phys. Rev. B}\ }\textbf {\bibinfo {volume} {91}},\
  \bibinfo {pages} {081103(R)} (\bibinfo {year} {2015})}\BibitemShut {NoStop}%
\bibitem [{\citenamefont {Khatami}\ \emph {et~al.}(2012)\citenamefont
  {Khatami}, \citenamefont {Rigol}, \citenamefont {Rela{\~n}o},\ and\
  \citenamefont {Garc{\'i}a-Garc{\'i}a}}]{khatami_quantum_2012}%
  \BibitemOpen
  \bibfield  {author} {\bibinfo {author} {\bibfnamefont {Ehsan}\ \bibnamefont
  {Khatami}}, \bibinfo {author} {\bibfnamefont {Marcos}\ \bibnamefont {Rigol}},
  \bibinfo {author} {\bibfnamefont {Armando}\ \bibnamefont {Rela{\~n}o}}, \
  and\ \bibinfo {author} {\bibfnamefont {Antonio~M.}\ \bibnamefont
  {Garc{\'i}a-Garc{\'i}a}},\ }\bibfield  {title} {\enquote {\bibinfo {title}
  {Quantum quenches in disordered systems: {Approach} to thermal equilibrium
  without a typical relaxation time},}\ }\href {\doibase
  10.1103/PhysRevE.85.050102} {\bibfield  {journal} {\bibinfo  {journal} {Phys.
  Rev. E}\ }\textbf {\bibinfo {volume} {85}},\ \bibinfo {pages} {050102}
  (\bibinfo {year} {2012})}\BibitemShut {NoStop}%
\bibitem [{\citenamefont {Mondaini}\ and\ \citenamefont
  {Rigol}(2015)}]{mondaini_many-body_2015}%
  \BibitemOpen
  \bibfield  {author} {\bibinfo {author} {\bibfnamefont {Rubem}\ \bibnamefont
  {Mondaini}}\ and\ \bibinfo {author} {\bibfnamefont {Marcos}\ \bibnamefont
  {Rigol}},\ }\bibfield  {title} {\enquote {\bibinfo {title} {Many-body
  localization and thermalization in disordered {Hubbard} chains},}\ }\href
  {\doibase 10.1103/PhysRevA.92.041601} {\bibfield  {journal} {\bibinfo
  {journal} {Phys. Rev. A}\ }\textbf {\bibinfo {volume} {92}},\ \bibinfo
  {pages} {041601} (\bibinfo {year} {2015})}\BibitemShut {NoStop}%
\bibitem [{\citenamefont {Nandkishore}\ and\ \citenamefont
  {Huse}(2015)}]{nandkishore_many-body_2015}%
  \BibitemOpen
  \bibfield  {author} {\bibinfo {author} {\bibfnamefont {Rahul}\ \bibnamefont
  {Nandkishore}}\ and\ \bibinfo {author} {\bibfnamefont {David~A.}\
  \bibnamefont {Huse}},\ }\bibfield  {title} {\enquote {\bibinfo {title}
  {Many-{Body} {Localization} and {Thermalization} in {Quantum} {Statistical}
  {Mechanics}},}\ }\href {\doibase 10.1146/annurev-conmatphys-031214-014726}
  {\bibfield  {journal} {\bibinfo  {journal} {Annual Review of Condensed Matter
  Physics}\ }\textbf {\bibinfo {volume} {6}},\ \bibinfo {pages} {15--38}
  (\bibinfo {year} {2015})}\BibitemShut {NoStop}%
\bibitem [{\citenamefont {Bauer}\ and\ \citenamefont
  {Nayak}(2013)}]{bauer_area_2013}%
  \BibitemOpen
  \bibfield  {author} {\bibinfo {author} {\bibfnamefont {Bela}\ \bibnamefont
  {Bauer}}\ and\ \bibinfo {author} {\bibfnamefont {Chetan}\ \bibnamefont
  {Nayak}},\ }\bibfield  {title} {\enquote {\bibinfo {title} {Area laws in a
  many-body localized state and its implications for topological order},}\
  }\href {\doibase 10.1088/1742-5468/2013/09/P09005} {\bibfield  {journal}
  {\bibinfo  {journal} {J. Stat. Mech.}\ }\textbf {\bibinfo {volume} {2013}},\
  \bibinfo {pages} {P09005} (\bibinfo {year} {2013})}\BibitemShut {NoStop}%
\bibitem [{\citenamefont {Kj{\"a}ll}\ \emph {et~al.}(2014)\citenamefont
  {Kj{\"a}ll}, \citenamefont {Bardarson},\ and\ \citenamefont
  {Pollmann}}]{kjall_many-body_2014}%
  \BibitemOpen
  \bibfield  {author} {\bibinfo {author} {\bibfnamefont {Jonas~A.}\
  \bibnamefont {Kj{\"a}ll}}, \bibinfo {author} {\bibfnamefont {Jens~H.}\
  \bibnamefont {Bardarson}}, \ and\ \bibinfo {author} {\bibfnamefont {Frank}\
  \bibnamefont {Pollmann}},\ }\bibfield  {title} {\enquote {\bibinfo {title}
  {Many-{Body} {Localization} in a {Disordered} {Quantum} {Ising} {Chain}},}\
  }\href {\doibase 10.1103/PhysRevLett.113.107204} {\bibfield  {journal}
  {\bibinfo  {journal} {Phys. Rev. Lett.}\ }\textbf {\bibinfo {volume} {113}},\
  \bibinfo {pages} {107204} (\bibinfo {year} {2014})}\BibitemShut {NoStop}%
\bibitem [{\citenamefont {Grover}(2014)}]{grover_certain_2014}%
  \BibitemOpen
  \bibfield  {author} {\bibinfo {author} {\bibfnamefont {Tarun}\ \bibnamefont
  {Grover}},\ }\bibfield  {title} {\enquote {\bibinfo {title} {Certain
  {General} {Constraints} on the {Many}-{Body} {Localization} {Transition}},}\
  }\href {http://arxiv.org/abs/1405.1471} {\bibfield  {journal} {\bibinfo
  {journal} {arXiv:1405.1471 [cond-mat, physics:quant-ph]}\ } (\bibinfo {year}
  {2014})},\ \bibinfo {note} {arXiv: 1405.1471}\BibitemShut {NoStop}%
\bibitem [{\citenamefont {Bar~Lev}\ \emph {et~al.}(2015)\citenamefont
  {Bar~Lev}, \citenamefont {Cohen},\ and\ \citenamefont
  {Reichman}}]{bar_lev_absence_2015}%
  \BibitemOpen
  \bibfield  {author} {\bibinfo {author} {\bibfnamefont {Yevgeny}\ \bibnamefont
  {Bar~Lev}}, \bibinfo {author} {\bibfnamefont {Guy}\ \bibnamefont {Cohen}}, \
  and\ \bibinfo {author} {\bibfnamefont {David~R.}\ \bibnamefont {Reichman}},\
  }\bibfield  {title} {\enquote {\bibinfo {title} {Absence of {Diffusion} in an
  {Interacting} {System} of {Spinless} {Fermions} on a {One}-{Dimensional}
  {Disordered} {Lattice}},}\ }\href {\doibase 10.1103/PhysRevLett.114.100601}
  {\bibfield  {journal} {\bibinfo  {journal} {Phys. Rev. Lett.}\ }\textbf
  {\bibinfo {volume} {114}},\ \bibinfo {pages} {100601} (\bibinfo {year}
  {2015})}\BibitemShut {NoStop}%
\bibitem [{\citenamefont {Vosk}\ \emph {et~al.}(2015)\citenamefont {Vosk},
  \citenamefont {Huse},\ and\ \citenamefont {Altman}}]{vosk_theory_2015}%
  \BibitemOpen
  \bibfield  {author} {\bibinfo {author} {\bibfnamefont {Ronen}\ \bibnamefont
  {Vosk}}, \bibinfo {author} {\bibfnamefont {David~A.}\ \bibnamefont {Huse}}, \
  and\ \bibinfo {author} {\bibfnamefont {Ehud}\ \bibnamefont {Altman}},\
  }\bibfield  {title} {\enquote {\bibinfo {title} {Theory of the {Many}-{Body}
  {Localization} {Transition} in {One}-{Dimensional} {Systems}},}\ }\href
  {\doibase 10.1103/PhysRevX.5.031032} {\bibfield  {journal} {\bibinfo
  {journal} {Phys. Rev. X}\ }\textbf {\bibinfo {volume} {5}},\ \bibinfo {pages}
  {031032} (\bibinfo {year} {2015})}\BibitemShut {NoStop}%
\bibitem [{\citenamefont {Chen}\ \emph {et~al.}(2015)\citenamefont {Chen},
  \citenamefont {Yu}, \citenamefont {Cho}, \citenamefont {Clark},\ and\
  \citenamefont {Fradkin}}]{chen_many-body_2015}%
  \BibitemOpen
  \bibfield  {author} {\bibinfo {author} {\bibfnamefont {Xiao}\ \bibnamefont
  {Chen}}, \bibinfo {author} {\bibfnamefont {Xiongjie}\ \bibnamefont {Yu}},
  \bibinfo {author} {\bibfnamefont {Gil~Young}\ \bibnamefont {Cho}}, \bibinfo
  {author} {\bibfnamefont {Bryan~K.}\ \bibnamefont {Clark}}, \ and\ \bibinfo
  {author} {\bibfnamefont {Eduardo}\ \bibnamefont {Fradkin}},\ }\bibfield
  {title} {\enquote {\bibinfo {title} {Many-body localization transition in
  {Rokhsar}-{Kivelson}-type wave functions},}\ }\href {\doibase
  10.1103/PhysRevB.92.214204} {\bibfield  {journal} {\bibinfo  {journal} {Phys.
  Rev. B}\ }\textbf {\bibinfo {volume} {92}},\ \bibinfo {pages} {214204}
  (\bibinfo {year} {2015})}\BibitemShut {NoStop}%
\bibitem [{\citenamefont {Pal}\ and\ \citenamefont
  {Huse}(2010)}]{pal_many-body_2010}%
  \BibitemOpen
  \bibfield  {author} {\bibinfo {author} {\bibfnamefont {Arijeet}\ \bibnamefont
  {Pal}}\ and\ \bibinfo {author} {\bibfnamefont {David~A.}\ \bibnamefont
  {Huse}},\ }\bibfield  {title} {\enquote {\bibinfo {title} {Many-body
  localization phase transition},}\ }\href {\doibase
  10.1103/PhysRevB.82.174411} {\bibfield  {journal} {\bibinfo  {journal} {Phys.
  Rev. B}\ }\textbf {\bibinfo {volume} {82}},\ \bibinfo {pages} {174411}
  (\bibinfo {year} {2010})}\BibitemShut {NoStop}%
\bibitem [{\citenamefont {Georgeot}\ and\ \citenamefont
  {Shepelyansky}(1998)}]{georgeot_integrability_1998}%
  \BibitemOpen
  \bibfield  {author} {\bibinfo {author} {\bibfnamefont {B.}~\bibnamefont
  {Georgeot}}\ and\ \bibinfo {author} {\bibfnamefont {D.~L.}\ \bibnamefont
  {Shepelyansky}},\ }\bibfield  {title} {\enquote {\bibinfo {title}
  {Integrability and {Quantum} {Chaos} in {Spin} {Glass} {Shards}},}\ }\href
  {\doibase 10.1103/PhysRevLett.81.5129} {\bibfield  {journal} {\bibinfo
  {journal} {Phys. Rev. Lett.}\ }\textbf {\bibinfo {volume} {81}},\ \bibinfo
  {pages} {5129--5132} (\bibinfo {year} {1998})}\BibitemShut {NoStop}%
\bibitem [{\citenamefont {Mondragon-Shem}\ \emph {et~al.}(2015)\citenamefont
  {Mondragon-Shem}, \citenamefont {Pal}, \citenamefont {Hughes},\ and\
  \citenamefont {Laumann}}]{mondragon-shem_many-body_2015}%
  \BibitemOpen
  \bibfield  {author} {\bibinfo {author} {\bibfnamefont {Ian}\ \bibnamefont
  {Mondragon-Shem}}, \bibinfo {author} {\bibfnamefont {Arijeet}\ \bibnamefont
  {Pal}}, \bibinfo {author} {\bibfnamefont {Taylor~L.}\ \bibnamefont {Hughes}},
  \ and\ \bibinfo {author} {\bibfnamefont {Chris~R.}\ \bibnamefont {Laumann}},\
  }\bibfield  {title} {\enquote {\bibinfo {title} {Many-body mobility edge due
  to symmetry-constrained dynamics and strong interactions},}\ }\href {\doibase
  10.1103/PhysRevB.92.064203} {\bibfield  {journal} {\bibinfo  {journal} {Phys.
  Rev. B}\ }\textbf {\bibinfo {volume} {92}},\ \bibinfo {pages} {064203}
  (\bibinfo {year} {2015})}\BibitemShut {NoStop}%
\bibitem [{\citenamefont {Bera}\ \emph {et~al.}(2015)\citenamefont {Bera},
  \citenamefont {Schomerus}, \citenamefont {Heidrich-Meisner},\ and\
  \citenamefont {Bardarson}}]{bera_many-body_2015}%
  \BibitemOpen
  \bibfield  {author} {\bibinfo {author} {\bibfnamefont {Soumya}\ \bibnamefont
  {Bera}}, \bibinfo {author} {\bibfnamefont {Henning}\ \bibnamefont
  {Schomerus}}, \bibinfo {author} {\bibfnamefont {Fabian}\ \bibnamefont
  {Heidrich-Meisner}}, \ and\ \bibinfo {author} {\bibfnamefont {Jens~H.}\
  \bibnamefont {Bardarson}},\ }\bibfield  {title} {\enquote {\bibinfo {title}
  {Many-{Body} {Localization} {Characterized} from a {One}-{Particle}
  {Perspective}},}\ }\href {\doibase 10.1103/PhysRevLett.115.046603} {\bibfield
   {journal} {\bibinfo  {journal} {Phys. Rev. Lett.}\ }\textbf {\bibinfo
  {volume} {115}},\ \bibinfo {pages} {046603} (\bibinfo {year}
  {2015})}\BibitemShut {NoStop}%
\bibitem [{\citenamefont {Laumann}\ \emph {et~al.}(2014)\citenamefont
  {Laumann}, \citenamefont {Pal},\ and\ \citenamefont
  {Scardicchio}}]{laumann_many-body_2014}%
  \BibitemOpen
  \bibfield  {author} {\bibinfo {author} {\bibfnamefont {C.~R.}\ \bibnamefont
  {Laumann}}, \bibinfo {author} {\bibfnamefont {A.}~\bibnamefont {Pal}}, \ and\
  \bibinfo {author} {\bibfnamefont {A.}~\bibnamefont {Scardicchio}},\
  }\bibfield  {title} {\enquote {\bibinfo {title} {Many-{Body} {Mobility}
  {Edge} in a {Mean}-{Field} {Quantum} {Spin} {Glass}},}\ }\href {\doibase
  10.1103/PhysRevLett.113.200405} {\bibfield  {journal} {\bibinfo  {journal}
  {Phys. Rev. Lett.}\ }\textbf {\bibinfo {volume} {113}},\ \bibinfo {pages}
  {200405} (\bibinfo {year} {2014})}\BibitemShut {NoStop}%
\bibitem [{\citenamefont {Serbyn}\ \emph {et~al.}(2015)\citenamefont {Serbyn},
  \citenamefont {Papi{\'c}},\ and\ \citenamefont
  {Abanin}}]{serbyn_criterion_2015}%
  \BibitemOpen
  \bibfield  {author} {\bibinfo {author} {\bibfnamefont {Maksym}\ \bibnamefont
  {Serbyn}}, \bibinfo {author} {\bibfnamefont {Z.}~\bibnamefont {Papi{\'c}}}, \
  and\ \bibinfo {author} {\bibfnamefont {Dmitry~A.}\ \bibnamefont {Abanin}},\
  }\bibfield  {title} {\enquote {\bibinfo {title} {Criterion for {Many}-{Body}
  {Localization}-{Delocalization} {Phase} {Transition}},}\ }\href {\doibase
  10.1103/PhysRevX.5.041047} {\bibfield  {journal} {\bibinfo  {journal} {Phys.
  Rev. X}\ }\textbf {\bibinfo {volume} {5}},\ \bibinfo {pages} {041047}
  (\bibinfo {year} {2015})}\BibitemShut {NoStop}%
\bibitem [{Note1()}]{Note1}%
  \BibitemOpen
  \bibinfo {note} {Except for the region very close to the clean limit at
  $h=0$, where the system becomes integrable.}\BibitemShut {Stop}%
\bibitem [{\citenamefont {Agarwal}\ \emph {et~al.}(2015)\citenamefont
  {Agarwal}, \citenamefont {Gopalakrishnan}, \citenamefont {Knap},
  \citenamefont {M{\"u}ller},\ and\ \citenamefont
  {Demler}}]{agarwal_anomalous_2015}%
  \BibitemOpen
  \bibfield  {author} {\bibinfo {author} {\bibfnamefont {Kartiek}\ \bibnamefont
  {Agarwal}}, \bibinfo {author} {\bibfnamefont {Sarang}\ \bibnamefont
  {Gopalakrishnan}}, \bibinfo {author} {\bibfnamefont {Michael}\ \bibnamefont
  {Knap}}, \bibinfo {author} {\bibfnamefont {Markus}\ \bibnamefont
  {M{\"u}ller}}, \ and\ \bibinfo {author} {\bibfnamefont {Eugene}\ \bibnamefont
  {Demler}},\ }\bibfield  {title} {\enquote {\bibinfo {title} {Anomalous
  {Diffusion} and {Griffiths} {Effects} {Near} the {Many}-{Body} {Localization}
  {Transition}},}\ }\href {\doibase 10.1103/PhysRevLett.114.160401} {\bibfield
  {journal} {\bibinfo  {journal} {Phys. Rev. Lett.}\ }\textbf {\bibinfo
  {volume} {114}},\ \bibinfo {pages} {160401} (\bibinfo {year}
  {2015})}\BibitemShut {NoStop}%
\bibitem [{\citenamefont {Lerose}\ \emph {et~al.}(2015)\citenamefont {Lerose},
  \citenamefont {Varma}, \citenamefont {Pietracaprina}, \citenamefont {Goold},\
  and\ \citenamefont {Scardicchio}}]{lerose_coexistence_2015}%
  \BibitemOpen
  \bibfield  {author} {\bibinfo {author} {\bibfnamefont {Alessio}\ \bibnamefont
  {Lerose}}, \bibinfo {author} {\bibfnamefont {Vipin~Kerala}\ \bibnamefont
  {Varma}}, \bibinfo {author} {\bibfnamefont {Francesca}\ \bibnamefont
  {Pietracaprina}}, \bibinfo {author} {\bibfnamefont {John}\ \bibnamefont
  {Goold}}, \ and\ \bibinfo {author} {\bibfnamefont {Antonello}\ \bibnamefont
  {Scardicchio}},\ }\bibfield  {title} {\enquote {\bibinfo {title} {Coexistence
  of energy diffusion and spin sub-diffusion in the ergodic phase of a many
  body localizable spin chain},}\ }\href {http://arxiv.org/abs/1511.09144}
  {\bibfield  {journal} {\bibinfo  {journal} {arXiv:1511.09144 [cond-mat,
  physics:quant-ph]}\ } (\bibinfo {year} {2015})},\ \bibinfo {note} {arXiv:
  1511.09144}\BibitemShut {NoStop}%
\bibitem [{\citenamefont {Luitz}\ \emph {et~al.}(2016)\citenamefont {Luitz},
  \citenamefont {Laflorencie},\ and\ \citenamefont
  {Alet}}]{luitz_extended_2016}%
  \BibitemOpen
  \bibfield  {author} {\bibinfo {author} {\bibfnamefont {David~J.}\
  \bibnamefont {Luitz}}, \bibinfo {author} {\bibfnamefont {Nicolas}\
  \bibnamefont {Laflorencie}}, \ and\ \bibinfo {author} {\bibfnamefont
  {Fabien}\ \bibnamefont {Alet}},\ }\bibfield  {title} {\enquote {\bibinfo
  {title} {Extended slow dynamical regime close to the many-body localization
  transition},}\ }\href {\doibase 10.1103/PhysRevB.93.060201} {\bibfield
  {journal} {\bibinfo  {journal} {Phys. Rev. B}\ }\textbf {\bibinfo {volume}
  {93}},\ \bibinfo {pages} {060201} (\bibinfo {year} {2016})}\BibitemShut
  {NoStop}%
\bibitem [{\citenamefont {Potter}\ \emph {et~al.}(2015)\citenamefont {Potter},
  \citenamefont {Vasseur},\ and\ \citenamefont
  {Parameswaran}}]{potter_universal_2015}%
  \BibitemOpen
  \bibfield  {author} {\bibinfo {author} {\bibfnamefont {Andrew~C.}\
  \bibnamefont {Potter}}, \bibinfo {author} {\bibfnamefont {Romain}\
  \bibnamefont {Vasseur}}, \ and\ \bibinfo {author} {\bibfnamefont {S.~A.}\
  \bibnamefont {Parameswaran}},\ }\bibfield  {title} {\enquote {\bibinfo
  {title} {Universal {Properties} of {Many}-{Body} {Delocalization}
  {Transitions}},}\ }\href {\doibase 10.1103/PhysRevX.5.031033} {\bibfield
  {journal} {\bibinfo  {journal} {Phys. Rev. X}\ }\textbf {\bibinfo {volume}
  {5}},\ \bibinfo {pages} {031033} (\bibinfo {year} {2015})}\BibitemShut
  {NoStop}%
\bibitem [{\citenamefont {Griffiths}(1969)}]{griffiths_nonanalytic_1969}%
  \BibitemOpen
  \bibfield  {author} {\bibinfo {author} {\bibfnamefont {Robert~B.}\
  \bibnamefont {Griffiths}},\ }\bibfield  {title} {\enquote {\bibinfo {title}
  {Nonanalytic {Behavior} {Above} the {Critical} {Point} in a {Random} {Ising}
  {Ferromagnet}},}\ }\href {\doibase 10.1103/PhysRevLett.23.17} {\bibfield
  {journal} {\bibinfo  {journal} {Phys. Rev. Lett.}\ }\textbf {\bibinfo
  {volume} {23}},\ \bibinfo {pages} {17--19} (\bibinfo {year}
  {1969})}\BibitemShut {NoStop}%
\bibitem [{\citenamefont {Vojta}(2010)}]{vojta_quantum_2010}%
  \BibitemOpen
  \bibfield  {author} {\bibinfo {author} {\bibfnamefont {Thomas}\ \bibnamefont
  {Vojta}},\ }\bibfield  {title} {\enquote {\bibinfo {title} {Quantum
  {Griffiths} {Effects} and {Smeared} {Phase} {Transitions} in {Metals}:
  {Theory} and {Experiment}},}\ }\href {\doibase 10.1007/s10909-010-0205-4}
  {\bibfield  {journal} {\bibinfo  {journal} {J Low Temp Phys}\ }\textbf
  {\bibinfo {volume} {161}},\ \bibinfo {pages} {299--323} (\bibinfo {year}
  {2010})}\BibitemShut {NoStop}%
\bibitem [{\citenamefont {Chiara}\ \emph {et~al.}(2006)\citenamefont {Chiara},
  \citenamefont {Montangero}, \citenamefont {Calabrese},\ and\ \citenamefont
  {Fazio}}]{chiara_entanglement_2006}%
  \BibitemOpen
  \bibfield  {author} {\bibinfo {author} {\bibfnamefont {Gabriele~De}\
  \bibnamefont {Chiara}}, \bibinfo {author} {\bibfnamefont {Simone}\
  \bibnamefont {Montangero}}, \bibinfo {author} {\bibfnamefont {Pasquale}\
  \bibnamefont {Calabrese}}, \ and\ \bibinfo {author} {\bibfnamefont {Rosario}\
  \bibnamefont {Fazio}},\ }\bibfield  {title} {\enquote {\bibinfo {title}
  {Entanglement entropy dynamics of {Heisenberg} chains},}\ }\href {\doibase
  10.1088/1742-5468/2006/03/P03001} {\bibfield  {journal} {\bibinfo  {journal}
  {J. Stat. Mech.}\ }\textbf {\bibinfo {volume} {2006}},\ \bibinfo {pages}
  {P03001} (\bibinfo {year} {2006})}\BibitemShut {NoStop}%
\bibitem [{\citenamefont {{\v Z}nidari{\v c}}\ \emph
  {et~al.}(2008)\citenamefont {{\v Z}nidari{\v c}}, \citenamefont {Prosen},\
  and\ \citenamefont {Prelov{\v s}ek}}]{znidaric_many-body_2008}%
  \BibitemOpen
  \bibfield  {author} {\bibinfo {author} {\bibfnamefont {Marko}\ \bibnamefont
  {{\v Z}nidari{\v c}}}, \bibinfo {author} {\bibfnamefont {Toma{\v z}}\
  \bibnamefont {Prosen}}, \ and\ \bibinfo {author} {\bibfnamefont {Peter}\
  \bibnamefont {Prelov{\v s}ek}},\ }\bibfield  {title} {\enquote {\bibinfo
  {title} {Many-body localization in the {Heisenberg} {XXZ} magnet in a random
  field},}\ }\href {\doibase 10.1103/PhysRevB.77.064426} {\bibfield  {journal}
  {\bibinfo  {journal} {Phys. Rev. B}\ }\textbf {\bibinfo {volume} {77}},\
  \bibinfo {pages} {064426} (\bibinfo {year} {2008})}\BibitemShut {NoStop}%
\bibitem [{\citenamefont {Bardarson}\ \emph {et~al.}(2012)\citenamefont
  {Bardarson}, \citenamefont {Pollmann},\ and\ \citenamefont
  {Moore}}]{bardarson_unbounded_2012}%
  \BibitemOpen
  \bibfield  {author} {\bibinfo {author} {\bibfnamefont {Jens~H.}\ \bibnamefont
  {Bardarson}}, \bibinfo {author} {\bibfnamefont {Frank}\ \bibnamefont
  {Pollmann}}, \ and\ \bibinfo {author} {\bibfnamefont {Joel~E.}\ \bibnamefont
  {Moore}},\ }\bibfield  {title} {\enquote {\bibinfo {title} {Unbounded
  {Growth} of {Entanglement} in {Models} of {Many}-{Body} {Localization}},}\
  }\href {\doibase 10.1103/PhysRevLett.109.017202} {\bibfield  {journal}
  {\bibinfo  {journal} {Phys. Rev. Lett.}\ }\textbf {\bibinfo {volume} {109}},\
  \bibinfo {pages} {017202} (\bibinfo {year} {2012})}\BibitemShut {NoStop}%
\bibitem [{\citenamefont {Serbyn}\ \emph {et~al.}(2013)\citenamefont {Serbyn},
  \citenamefont {Papi{\'c}},\ and\ \citenamefont {Abanin}}]{serbyn_local_2013}%
  \BibitemOpen
  \bibfield  {author} {\bibinfo {author} {\bibfnamefont {Maksym}\ \bibnamefont
  {Serbyn}}, \bibinfo {author} {\bibfnamefont {Z.}~\bibnamefont {Papi{\'c}}}, \
  and\ \bibinfo {author} {\bibfnamefont {Dmitry~A.}\ \bibnamefont {Abanin}},\
  }\bibfield  {title} {\enquote {\bibinfo {title} {Local {Conservation} {Laws}
  and the {Structure} of the {Many}-{Body} {Localized} {States}},}\ }\href
  {\doibase 10.1103/PhysRevLett.111.127201} {\bibfield  {journal} {\bibinfo
  {journal} {Phys. Rev. Lett.}\ }\textbf {\bibinfo {volume} {111}},\ \bibinfo
  {pages} {127201} (\bibinfo {year} {2013})}\BibitemShut {NoStop}%
\bibitem [{\citenamefont {Huse}\ \emph {et~al.}(2014)\citenamefont {Huse},
  \citenamefont {Nandkishore},\ and\ \citenamefont
  {Oganesyan}}]{huse_phenomenology_2014}%
  \BibitemOpen
  \bibfield  {author} {\bibinfo {author} {\bibfnamefont {David~A.}\
  \bibnamefont {Huse}}, \bibinfo {author} {\bibfnamefont {Rahul}\ \bibnamefont
  {Nandkishore}}, \ and\ \bibinfo {author} {\bibfnamefont {Vadim}\ \bibnamefont
  {Oganesyan}},\ }\bibfield  {title} {\enquote {\bibinfo {title} {Phenomenology
  of fully many-body-localized systems},}\ }\href {\doibase
  10.1103/PhysRevB.90.174202} {\bibfield  {journal} {\bibinfo  {journal} {Phys.
  Rev. B}\ }\textbf {\bibinfo {volume} {90}},\ \bibinfo {pages} {174202}
  (\bibinfo {year} {2014})}\BibitemShut {NoStop}%
\bibitem [{\citenamefont {Imbrie}(2014)}]{imbrie_many-body_2014}%
  \BibitemOpen
  \bibfield  {author} {\bibinfo {author} {\bibfnamefont {John~Z.}\ \bibnamefont
  {Imbrie}},\ }\bibfield  {title} {\enquote {\bibinfo {title} {On {Many}-{Body}
  {Localization} for {Quantum} {Spin} {Chains}},}\ }\href
  {http://arxiv.org/abs/1403.7837} {\bibfield  {journal} {\bibinfo  {journal}
  {arXiv:1403.7837 [cond-mat, physics:math-ph]}\ } (\bibinfo {year} {2014})},\
  \bibinfo {note} {arXiv: 1403.7837}\BibitemShut {NoStop}%
\bibitem [{\citenamefont {Luca}\ and\ \citenamefont
  {Scardicchio}(2013)}]{luca_ergodicity_2013}%
  \BibitemOpen
  \bibfield  {author} {\bibinfo {author} {\bibfnamefont {A.~De}\ \bibnamefont
  {Luca}}\ and\ \bibinfo {author} {\bibfnamefont {A.}~\bibnamefont
  {Scardicchio}},\ }\bibfield  {title} {\enquote {\bibinfo {title} {Ergodicity
  breaking in a model showing many-body localization},}\ }\href {\doibase
  10.1209/0295-5075/101/37003} {\bibfield  {journal} {\bibinfo  {journal}
  {EPL}\ }\textbf {\bibinfo {volume} {101}},\ \bibinfo {pages} {37003}
  (\bibinfo {year} {2013})}\BibitemShut {NoStop}%
\bibitem [{\citenamefont {Stewart}(2002)}]{stewart_krylov--schur_2002}%
  \BibitemOpen
  \bibfield  {author} {\bibinfo {author} {\bibfnamefont {G.}~\bibnamefont
  {Stewart}},\ }\bibfield  {title} {\enquote {\bibinfo {title} {A
  {Krylov}--{Schur} {Algorithm} for {Large} {Eigenproblems}},}\ }\href
  {\doibase 10.1137/S0895479800371529} {\bibfield  {journal} {\bibinfo
  {journal} {SIAM. J. Matrix Anal. \& Appl.}\ }\textbf {\bibinfo {volume}
  {23}},\ \bibinfo {pages} {601--614} (\bibinfo {year} {2002})}\BibitemShut
  {NoStop}%
\bibitem [{\citenamefont {Garrison}\ and\ \citenamefont
  {Grover}(2015)}]{garrison_does_2015}%
  \BibitemOpen
  \bibfield  {author} {\bibinfo {author} {\bibfnamefont {James~R.}\
  \bibnamefont {Garrison}}\ and\ \bibinfo {author} {\bibfnamefont {Tarun}\
  \bibnamefont {Grover}},\ }\bibfield  {title} {\enquote {\bibinfo {title}
  {Does a single eigenstate encode the full {Hamiltonian}?}}\ }\href
  {http://arxiv.org/abs/1503.00729} {\bibfield  {journal} {\bibinfo  {journal}
  {arXiv:1503.00729 [cond-mat, physics:hep-th, physics:quant-ph]}\ } (\bibinfo
  {year} {2015})},\ \bibinfo {note} {arXiv: 1503.00729}\BibitemShut {NoStop}%
\bibitem [{Note2()}]{Note2}%
  \BibitemOpen
  \bibinfo {note} {This is more apparent for larger subsystem magnetizations,
  where the distribution has a larger number of peaks.}\BibitemShut {Stop}%
\bibitem [{\citenamefont {Baygan}\ \emph {et~al.}(2015)\citenamefont {Baygan},
  \citenamefont {Lim},\ and\ \citenamefont {Sheng}}]{baygan_many-body_2015-1}%
  \BibitemOpen
  \bibfield  {author} {\bibinfo {author} {\bibfnamefont {Elliott}\ \bibnamefont
  {Baygan}}, \bibinfo {author} {\bibfnamefont {S.~P.}\ \bibnamefont {Lim}}, \
  and\ \bibinfo {author} {\bibfnamefont {D.~N.}\ \bibnamefont {Sheng}},\
  }\bibfield  {title} {\enquote {\bibinfo {title} {Many-body localization and
  mobility edge in a disordered spin-1/2 {Heisenberg} ladder},}\ }\href
  {\doibase 10.1103/PhysRevB.92.195153} {\bibfield  {journal} {\bibinfo
  {journal} {Phys. Rev. B}\ }\textbf {\bibinfo {volume} {92}},\ \bibinfo
  {pages} {195153} (\bibinfo {year} {2015})}\BibitemShut {NoStop}%
\bibitem [{\citenamefont {Lim}\ and\ \citenamefont
  {Sheng}(2015)}]{lim_nature_2015}%
  \BibitemOpen
  \bibfield  {author} {\bibinfo {author} {\bibfnamefont {S.~P.}\ \bibnamefont
  {Lim}}\ and\ \bibinfo {author} {\bibfnamefont {D.~N.}\ \bibnamefont
  {Sheng}},\ }\bibfield  {title} {\enquote {\bibinfo {title} {Nature of
  {Many}-{Body} {Localization} and {Transitions} by {Density} {Matrix}
  {Renormaliztion} {Group} and {Exact} {Diagonalization} {Studies}},}\ }\href
  {http://arxiv.org/abs/1510.08145} {\bibfield  {journal} {\bibinfo  {journal}
  {arXiv:1510.08145 [cond-mat]}\ } (\bibinfo {year} {2015})},\ \bibinfo {note}
  {arXiv: 1510.08145}\BibitemShut {NoStop}%
\bibitem [{Note3()}]{Note3}%
  \BibitemOpen
  \bibinfo {note} {It should be noted that the differences in the details of
  the distributions in Ref. \protect \rev@citealpnum {lim_nature_2015} and the
  present work are due to different boundary conditions: Here we use periodic
  boundaries, while Ref. \protect \rev@citealpnum {lim_nature_2015} employ open
  boundary conditions.}\BibitemShut {Stop}%
\bibitem [{\citenamefont {Bera}\ and\ \citenamefont
  {Lakshminarayan}(2015)}]{bera_local_2015}%
  \BibitemOpen
  \bibfield  {author} {\bibinfo {author} {\bibfnamefont {Soumya}\ \bibnamefont
  {Bera}}\ and\ \bibinfo {author} {\bibfnamefont {Arul}\ \bibnamefont
  {Lakshminarayan}},\ }\bibfield  {title} {\enquote {\bibinfo {title} {Local
  entanglement structure across a many-body localization transition},}\ }\href
  {http://arxiv.org/abs/1512.04705} {\bibfield  {journal} {\bibinfo  {journal}
  {arXiv:1512.04705 [cond-mat]}\ } (\bibinfo {year} {2015})},\ \bibinfo {note}
  {arXiv: 1512.04705}\BibitemShut {NoStop}%
\bibitem [{\citenamefont {Yu}\ \emph {et~al.}(2015)\citenamefont {Yu},
  \citenamefont {Pekker},\ and\ \citenamefont {Clark}}]{yu_finding_2015}%
  \BibitemOpen
  \bibfield  {author} {\bibinfo {author} {\bibfnamefont {Xiongjie}\
  \bibnamefont {Yu}}, \bibinfo {author} {\bibfnamefont {David}\ \bibnamefont
  {Pekker}}, \ and\ \bibinfo {author} {\bibfnamefont {Bryan~K.}\ \bibnamefont
  {Clark}},\ }\bibfield  {title} {\enquote {\bibinfo {title} {Finding matrix
  product state representations of highly-excited eigenstates of many-body
  localized {Hamiltonians}},}\ }\href {http://arxiv.org/abs/1509.01244}
  {\bibfield  {journal} {\bibinfo  {journal} {arXiv:1509.01244 [cond-mat]}\ }
  (\bibinfo {year} {2015})},\ \bibinfo {note} {arXiv: 1509.01244}\BibitemShut
  {NoStop}%
\bibitem [{\citenamefont {Khemani}\ \emph {et~al.}(2015)\citenamefont
  {Khemani}, \citenamefont {Pollmann},\ and\ \citenamefont
  {Sondhi}}]{khemani_obtaining_2015}%
  \BibitemOpen
  \bibfield  {author} {\bibinfo {author} {\bibfnamefont {Vedika}\ \bibnamefont
  {Khemani}}, \bibinfo {author} {\bibfnamefont {Frank}\ \bibnamefont
  {Pollmann}}, \ and\ \bibinfo {author} {\bibfnamefont {S.~L.}\ \bibnamefont
  {Sondhi}},\ }\bibfield  {title} {\enquote {\bibinfo {title} {Obtaining
  highly-excited eigenstates of many-body localized {Hamiltonians} by the
  density matrix renormalization group},}\ }\href
  {http://arxiv.org/abs/1509.00483} {\bibfield  {journal} {\bibinfo  {journal}
  {arXiv:1509.00483 [cond-mat]}\ } (\bibinfo {year} {2015})},\ \bibinfo {note}
  {arXiv: 1509.00483}\BibitemShut {NoStop}%
\bibitem [{\citenamefont {Kennes}\ and\ \citenamefont
  {Karrasch}(2015)}]{kennes_entanglement_2015}%
  \BibitemOpen
  \bibfield  {author} {\bibinfo {author} {\bibfnamefont {D.~M.}\ \bibnamefont
  {Kennes}}\ and\ \bibinfo {author} {\bibfnamefont {C.}~\bibnamefont
  {Karrasch}},\ }\bibfield  {title} {\enquote {\bibinfo {title} {Entanglement
  scaling of excited states in large one-dimensional many-body localized
  systems},}\ }\href {http://arxiv.org/abs/1511.02205} {\bibfield  {journal}
  {\bibinfo  {journal} {arXiv:1511.02205 [cond-mat]}\ } (\bibinfo {year}
  {2015})},\ \bibinfo {note} {arXiv: 1511.02205}\BibitemShut {NoStop}%
\bibitem [{\citenamefont {Balay}\ \emph
  {et~al.}(2014{\natexlab{a}})\citenamefont {Balay}, \citenamefont {Abhyankar},
  \citenamefont {Adams}, \citenamefont {Brown}, \citenamefont {Brune},
  \citenamefont {Buschelman}, \citenamefont {Eijkhout}, \citenamefont {Gropp},
  \citenamefont {Kaushik}, \citenamefont {Knepley}, \citenamefont {McInnes},
  \citenamefont {Rupp}, \citenamefont {Smith},\ and\ \citenamefont
  {Zhang}}]{petsc-web-page}%
  \BibitemOpen
  \bibfield  {author} {\bibinfo {author} {\bibfnamefont {Satish}\ \bibnamefont
  {Balay}}, \bibinfo {author} {\bibfnamefont {Shrirang}\ \bibnamefont
  {Abhyankar}}, \bibinfo {author} {\bibfnamefont {Mark~F.}\ \bibnamefont
  {Adams}}, \bibinfo {author} {\bibfnamefont {Jed}\ \bibnamefont {Brown}},
  \bibinfo {author} {\bibfnamefont {Peter}\ \bibnamefont {Brune}}, \bibinfo
  {author} {\bibfnamefont {Kris}\ \bibnamefont {Buschelman}}, \bibinfo {author}
  {\bibfnamefont {Victor}\ \bibnamefont {Eijkhout}}, \bibinfo {author}
  {\bibfnamefont {William~D.}\ \bibnamefont {Gropp}}, \bibinfo {author}
  {\bibfnamefont {Dinesh}\ \bibnamefont {Kaushik}}, \bibinfo {author}
  {\bibfnamefont {Matthew~G.}\ \bibnamefont {Knepley}}, \bibinfo {author}
  {\bibfnamefont {Lois~Curfman}\ \bibnamefont {McInnes}}, \bibinfo {author}
  {\bibfnamefont {Karl}\ \bibnamefont {Rupp}}, \bibinfo {author} {\bibfnamefont
  {Barry~F.}\ \bibnamefont {Smith}}, \ and\ \bibinfo {author} {\bibfnamefont
  {Hong}\ \bibnamefont {Zhang}},\ }\href {http://www.mcs.anl.gov/petsc}
  {\enquote {\bibinfo {title} {{PETS}c {W}eb page},}\ }\bibinfo {howpublished}
  {\url{http://www.mcs.anl.gov/petsc}} (\bibinfo {year}
  {2014}{\natexlab{a}})\BibitemShut {NoStop}%
\bibitem [{\citenamefont {Balay}\ \emph
  {et~al.}(2014{\natexlab{b}})\citenamefont {Balay}, \citenamefont {Abhyankar},
  \citenamefont {Adams}, \citenamefont {Brown}, \citenamefont {Brune},
  \citenamefont {Buschelman}, \citenamefont {Eijkhout}, \citenamefont {Gropp},
  \citenamefont {Kaushik}, \citenamefont {Knepley}, \citenamefont {McInnes},
  \citenamefont {Rupp}, \citenamefont {Smith},\ and\ \citenamefont
  {Zhang}}]{petsc-user-ref}%
  \BibitemOpen
  \bibfield  {author} {\bibinfo {author} {\bibfnamefont {Satish}\ \bibnamefont
  {Balay}}, \bibinfo {author} {\bibfnamefont {Shrirang}\ \bibnamefont
  {Abhyankar}}, \bibinfo {author} {\bibfnamefont {Mark~F.}\ \bibnamefont
  {Adams}}, \bibinfo {author} {\bibfnamefont {Jed}\ \bibnamefont {Brown}},
  \bibinfo {author} {\bibfnamefont {Peter}\ \bibnamefont {Brune}}, \bibinfo
  {author} {\bibfnamefont {Kris}\ \bibnamefont {Buschelman}}, \bibinfo {author}
  {\bibfnamefont {Victor}\ \bibnamefont {Eijkhout}}, \bibinfo {author}
  {\bibfnamefont {William~D.}\ \bibnamefont {Gropp}}, \bibinfo {author}
  {\bibfnamefont {Dinesh}\ \bibnamefont {Kaushik}}, \bibinfo {author}
  {\bibfnamefont {Matthew~G.}\ \bibnamefont {Knepley}}, \bibinfo {author}
  {\bibfnamefont {Lois~Curfman}\ \bibnamefont {McInnes}}, \bibinfo {author}
  {\bibfnamefont {Karl}\ \bibnamefont {Rupp}}, \bibinfo {author} {\bibfnamefont
  {Barry~F.}\ \bibnamefont {Smith}}, \ and\ \bibinfo {author} {\bibfnamefont
  {Hong}\ \bibnamefont {Zhang}},\ }\href {http://www.mcs.anl.gov/petsc} {\emph
  {\bibinfo {title} {{PETS}c Users Manual}}},\ \bibinfo {type} {Tech. Rep.}\
  \bibinfo {number} {ANL-95/11 - Revision 3.5}\ (\bibinfo  {institution}
  {Argonne National Laboratory},\ \bibinfo {year} {2014})\BibitemShut {NoStop}%
\bibitem [{\citenamefont {Balay}\ \emph {et~al.}(1997)\citenamefont {Balay},
  \citenamefont {Gropp}, \citenamefont {McInnes},\ and\ \citenamefont
  {Smith}}]{petsc-efficient}%
  \BibitemOpen
  \bibfield  {author} {\bibinfo {author} {\bibfnamefont {Satish}\ \bibnamefont
  {Balay}}, \bibinfo {author} {\bibfnamefont {William~D.}\ \bibnamefont
  {Gropp}}, \bibinfo {author} {\bibfnamefont {Lois~Curfman}\ \bibnamefont
  {McInnes}}, \ and\ \bibinfo {author} {\bibfnamefont {Barry~F.}\ \bibnamefont
  {Smith}},\ }\bibfield  {title} {\enquote {\bibinfo {title} {Efficient
  management of parallelism in object oriented numerical software libraries},}\
  }in\ \href@noop {} {\emph {\bibinfo {booktitle} {Modern Software Tools in
  Scientific Computing}}},\ \bibinfo {editor} {edited by\ \bibinfo {editor}
  {\bibfnamefont {E.}~\bibnamefont {Arge}}, \bibinfo {editor} {\bibfnamefont
  {A.~M.}\ \bibnamefont {Bruaset}}, \ and\ \bibinfo {editor} {\bibfnamefont
  {H.~P.}\ \bibnamefont {Langtangen}}}\ (\bibinfo  {publisher}
  {Birkh{\"{a}}user Press},\ \bibinfo {year} {1997})\ pp.\ \bibinfo {pages}
  {163--202}\BibitemShut {NoStop}%
\bibitem [{\citenamefont {Hernandez}\ \emph {et~al.}(2005)\citenamefont
  {Hernandez}, \citenamefont {Roman},\ and\ \citenamefont
  {Vidal}}]{hernandez_slepc:_2005}%
  \BibitemOpen
  \bibfield  {author} {\bibinfo {author} {\bibfnamefont {Vicente}\ \bibnamefont
  {Hernandez}}, \bibinfo {author} {\bibfnamefont {Jose~E.}\ \bibnamefont
  {Roman}}, \ and\ \bibinfo {author} {\bibfnamefont {Vicente}\ \bibnamefont
  {Vidal}},\ }\bibfield  {title} {\enquote {\bibinfo {title} {{SLEPc}: {A}
  {Scalable} and {Flexible} {Toolkit} for the {Solution} of {Eigenvalue}
  {Problems}},}\ }\href {\doibase 10.1145/1089014.1089019} {\bibfield
  {journal} {\bibinfo  {journal} {ACM Trans. Math. Softw.}\ }\textbf {\bibinfo
  {volume} {31}},\ \bibinfo {pages} {351--362} (\bibinfo {year}
  {2005})}\BibitemShut {NoStop}%
\bibitem [{\citenamefont {Amestoy}\ \emph {et~al.}(2001)\citenamefont
  {Amestoy}, \citenamefont {Duff}, \citenamefont {Koster},\ and\ \citenamefont
  {L'Excellent}}]{MUMPS1}%
  \BibitemOpen
  \bibfield  {author} {\bibinfo {author} {\bibfnamefont {P.~R.}\ \bibnamefont
  {Amestoy}}, \bibinfo {author} {\bibfnamefont {I.~S.}\ \bibnamefont {Duff}},
  \bibinfo {author} {\bibfnamefont {J.}~\bibnamefont {Koster}}, \ and\ \bibinfo
  {author} {\bibfnamefont {J.-Y.}\ \bibnamefont {L'Excellent}},\ }\bibfield
  {title} {\enquote {\bibinfo {title} {A fully asynchronous multifrontal solver
  using distributed dynamic scheduling},}\ }\href@noop {} {\bibfield  {journal}
  {\bibinfo  {journal} {SIAM J. Matrix Anal. Appl.}\ }\textbf {\bibinfo
  {volume} {23}},\ \bibinfo {pages} {15--41} (\bibinfo {year}
  {2001})}\BibitemShut {NoStop}%
\bibitem [{\citenamefont {Amestoy}\ \emph {et~al.}(2006)\citenamefont
  {Amestoy}, \citenamefont {Guermouche}, \citenamefont {L'Excellent},\ and\
  \citenamefont {Pralet}}]{MUMPS2}%
  \BibitemOpen
  \bibfield  {author} {\bibinfo {author} {\bibfnamefont {P.~R.}\ \bibnamefont
  {Amestoy}}, \bibinfo {author} {\bibfnamefont {A.}~\bibnamefont {Guermouche}},
  \bibinfo {author} {\bibfnamefont {J.-Y.}\ \bibnamefont {L'Excellent}}, \ and\
  \bibinfo {author} {\bibfnamefont {S.}~\bibnamefont {Pralet}},\ }\bibfield
  {title} {\enquote {\bibinfo {title} {Hybrid scheduling for the parallel
  solution of linear systems},}\ }\href@noop {} {\bibfield  {journal} {\bibinfo
   {journal} {Parallel Computing}\ }\textbf {\bibinfo {volume} {32}},\ \bibinfo
  {pages} {136--156} (\bibinfo {year} {2006})}\BibitemShut {NoStop}%
\end{thebibliography}%
\end{document}